\documentclass[prd,aps,showpacs,nofootinbib,eqsecnum,onecolumn,groupedaddress,
amssymb]{revtex4}
\usepackage{amsmath}
\usepackage{amssymb}
\usepackage{amsfonts}
\usepackage{graphicx,bm}
\usepackage{dcolumn}
\usepackage{color,amsxtra}
\usepackage{epsf}
\usepackage{enumerate}
\usepackage{hhline}
\usepackage{array}
\usepackage{tabularx}
\usepackage{subfigure}
\usepackage{appendix}
\usepackage[colorlinks=true]{hyperref}

\begin{document}

\title{The inflationary universe in F(R) gravity with antisymmetric tensor fields and their suppression during the universe evolution
}
\author{Emilio Elizalde}
\email{elizalde@ieec.uab.es}
\affiliation{Institut de Ci\`encies de l'Espai (ICE-CSIC/IEEC),\\ 
Campus UAB, c. Can Magrans s/n, 08193, Barcelona, Spain.}

\author{Sergei D. Odintsov}
\email{odintsov@ieec.uab.es}
\affiliation{Institut de Ci\`encies de l'Espai (ICE-CSIC/IEEC),\\ 
Campus UAB, c. Can Magrans s/n, 08193, Barcelona, Spain.}
\affiliation{Instituci\'o Catalana de Recerca i Estudis Avan\c{c}ats (ICREA),
Barcelona, Spain}

\author{Tanmoy Paul}
\email{pul.tnmy9@gmail.com}
\affiliation{Department of Theoretical Physics,\\
Indian Association for the Cultivation of Science,\\
2A $\&$ 2B Raja S.C. Mullick Road,\\
Kolkata - 700 032, India.}

\author{Diego S\'aez-Chill\'on G\'omez}
\email{diego.saez@ehu.eus}
\affiliation{Department of Theoretical Physics and History of Sciences, University of the Basque Country UPV/EHU, \\ P.O. Box 644, 48080 Bilbao, Spain}

\begin{abstract}
The intriguing question, why  the present scale of the universe is free from any perceptible footprints of rank-2 antisymmetric tensor fields? (generally known as Kalb-Ramond fields) is addressed. 
A quite natural explanation of this issue is given from the angle of higher-curvature gravity, both in four- and in five-dimensional spacetime.
The results here obtained reveal that the amplitude of the Kalb-Ramond field may be actually large and play a significant role during the early universe, while the presence of higher-order gravity 
suppresses this field during the cosmological evolution, so that it eventually becomes negligible in the current universe. Besides the suppression of the Kalb-Ramond field, the extra degree of freedom 
in $F(R)$ gravity, usually known as scalaron, also turns out to be responsible for inflation. Such F(R) gravity with Kalb-Ramond fields may govern the early universe to undergo an inflationary stage at 
early times (with the subsequent graceful exit) for wider range of F(R) gravity than without antisymmetric fields.. Furthermore,  the models---in four- and five-dimensional spacetimes---are linked to 
observational constraints, with the conclusion that the corresponding values of the spectral index and tensor-to-scalar ratio closely match the values provided by the Planck survey 2018 data.
\end{abstract}
\maketitle
\newpage

\section{Introduction}
A surprising feature of the present universe is that it carries no noticeable footprints
of higher rank (rank two or higher) antisymmetric tensor fields.  Apart from being the massless $(1,1)$ representation of the Lorentz group, such fields also arise naturally as closed string 
modes \cite{duff} and, consequently, are of considerable interest in string theory. In this context, the second rank antisymmetric 
tensor fields, generally known as Kalb-Ramond (KR) fields \cite{kalb}, have drawn considerable attention and have been extensively studied. However, 
dimensional analysis demands that the coupling strength of the KR field to other matter fields should go as $1/M_p$ ($M_p$ being the four-dimensional Planck mass), i.e. share the same dimensional 
coupling as the graviton. In spite of this, the large scale behaviour 
of the present universe appears to be governed solely by gravity and there is no experimental evidence of any second-rank antisymmetric 
Kalb-Ramond field being present. Therefore, if the KR field exists at all, it clearly must be severely suppressed at the present scale of our universe. This raises a natural question: why are the 
effects of the KR field less perceptible than the force of 
gravitation? Some attempts have been made to solve this puzzle both in four-dimensional as well as in higher-dimensional braneworld models \cite{sengupta1,sengupta2,sengupta3,sengupta4,tanmoy1}. 
In the context of higher-dimensional models, the warping geometrical nature of extra dimensions causes a huge suppression of the amplitude of the bulk for the KR field on our visible 3-brane.\\

However, the suppression of KR field still awaits a proper understanding in the context of cosmology. On the other hand, it has been 
shown that the energy density associated with KR field is large and might play a relevant role in the early universe. This fact, along with present day observations, have inspired us to study the 
cosmological evolution of the KR field from the very early universe, where it is also crucial to investigate whether the universe goes 
through an inflationary stage or not. In particular, we are interested in analysing the possible evolution of the Kalb-Ramond field  from the very early stages of the universe, and whether a 
suppression of this field can be achieved, in order to satisfy the observational constraints that we have at present. In addition, we also intend here to study whether inflation can still be 
realised in the presence of the KR field, including its compulsory graceful exit. The values of the spectral index, $n_s$, and the tensor-to-scalar ratio, $r$, are obtained and compared to the 
most recent Planck data available \cite{Planck}.
%
The present paper is,  a serious attempt to provide a natural explanation of the questions above in the framework of $F(R)$ gravity, both in four- 
as in five-dimensional spacetimes.\\

It is well known that Einstein-Hilbert term can be generalized to include higher order curvature terms in the gravitational action, 
which naturally arise from the diffeomorphism property of the action. Such higher order curvature terms may have their origin 
in String theory, such that naturally arise in the gravitational action \cite{Nojiri:2006je}. $F(R)$ gravity 
\cite{report,laurentis,delaCruzDombriz:2012xy,troisi,brooker,paliathanasis,tp1,bahamonde,ssg1,elizalde,gomez,linder,bamba,logR,tp_fermion,Olmo:2011ja}, Gauss-Bonnet 
(GB) \cite{nojiri3,cognola,odintsov_GB,Elizalde:2010jx} 
or more generally Lanczos-Lovelock gravity \cite{lanczos,lovelock} 
are some of the well known higher order curvature gravitational theories. While GB or Lanczos-Lovelock gravity have non-trivial 
consequences besides in higher dimensions, 
$F(R)$ gravity survives even in four dimensional spacetime model. For some choices of $F(R)$ (for which 
$F'(R)>0$), the corresponding model becomes free of ghosts.\\

On the other hand, over the last two decades, models with extra spatial dimensions \cite{arkani,horava,RS,kaloper,cohen,burgess,chodos} 
have been increasingly playing a central role in physics beyond standard model of particle physics \cite{rattazzi} and cosmology 
\cite{marteens,odintsov_brane1,odintsov_brane2,tp_inflation,tp_bouncing}. In all such models our 
visible universe is identified with a 3-brane embedded within a higher dimensional spacetime. Among all, the so-called Randall-Sundrum (RS) model \cite{RS} have gained a special 
attention as solves the gauge hierarchy problem 
without introducing any intermediate scale (between Planck and TeV) in the theory. RS scenario assumes one extra spatial 
dimension (in addition to the usual three spatial dimensions) with $S^1/Z_2$ orbifolding where the orbifolded fixed points are identified with 
two 3-branes. The intermediate region between the branes is fixed as a bulk which has a curvature of Planck order. In such 
higher order curvature regime, $F(R)$ gravity is supposed to play a relevant role. However all the higher dimensional 
braneworld scenarios demand a certain mechanism for stabilization of interbrane separation, also known as modulus or 
radion \cite{ssg2,GW,GW_radion,csaki}. Here, we show that higher order curvature degree(s) of freedom can generate a potential term 
for the radion field and fulfills the purpose of modulus stabilization. Keeping this in mind, here we try to address the cosmological evolution of antisymmetric Kalb-Ramond field in 
$F(R)$ gravity by two analysing two frameworks: the KR field in four dimensions and in a  higher dimensional bulk spacetime.\\

The paper is organized as follows : in section \ref{sectionFR} we briefly describe the equivalence between $F(R)$ model and scalar-tensor (ST) theory 
in D dimensions. Section \ref{Model-I} and \ref{Model-II} are devoted to the analysis of the cosmological evolution of the KR field in four and five dimensional spacetime 
in $F(R)$ gravity, respectively. The paper ends with some conclusive remarks and discussions in section \ref{conclusion}.

\section{$F(R)$ gravity and its scalar-tensor counterpart in D-dimensions}
\label{sectionFR}

In this section, we briefly describe $F(R)$ gravity in D-dimensions and its conformal picture in the Einstein frame, which results in the Hilbert-Einstein action with the presence of a 
scalar field. The $F(R)$ action can be written as follows:
\begin{eqnarray}
 S = \int d^Dx \sqrt{-G}\bigg[\frac{F(R)}{2\kappa^2}\bigg]
 \label{transformation1}
\end{eqnarray}
where $G$ is the determinant of D dimensional metric $G_{MN}$ ($M$, $N$ runs from $0$ to $D-1$), $R$ is the D dimensional Ricci scalar and 
$\frac{1}{2\kappa^2} = M^{D-2}$ with $M$ is the D dimensional Planck mass. 
By introducing an auxiliary field $A(x)$, the action (\ref{transformation1}) can be rewritten as,
\begin{eqnarray}
 S = \int d^Dx \sqrt{-G}\frac{1}{2\kappa^2}\bigg[F'(A)(R-A) + F(A)\bigg]
 \label{transformation2}
\end{eqnarray}
The variation of this action over the auxiliary field $A(x)$ leads to $A = R$, which finally results in the original action (\ref{transformation1}). Moreover, the action (\ref{transformation2}) 
can be mapped into the Einstein frame by applying the following conformal transformation on the metric $G_{MN}(x)$,
\begin{equation}
 G_{MN}(x) \longrightarrow \widetilde{G}_{MN}(x) = e^{-[\sqrt{\frac{4}{(D-1)(D-2)}}\kappa\xi]} G_{MN}(x)
 \label{transformation3}
\end{equation}
where $\xi(x)$ is the conformal factor which is related to the auxiliary field as $F'(A) = e^{-[\sqrt{\frac{D-2}{D-1}}\kappa\xi]}$, while $R$ and $\tilde{R}$ are the Ricci scalars in terms of 
the metrics $G_{MN}$ and $\tilde{G}_{MN}$ respectively, such that they are related by
\begin{eqnarray}
 R =e^{-[\sqrt{\frac{4}{(D-1)(D-2)}}\kappa\xi]}\bigg[\tilde{R} - \kappa^2\tilde{G}^{MN}\partial_{M}\xi\partial_{N}\xi 
 + 2\kappa\sqrt{\frac{D-1}{D-2}}\tilde{\Box}\xi\bigg]
 \nonumber
\end{eqnarray}
where $\tilde{\Box}$ represents the d'Alembertian operator formed by $\tilde{G}_{MN}$. Using the above expression along with the aforementioned relation among $\xi(x)$ and $F'(A)$, the following 
scalar-tensor action is achieved:

\begin{eqnarray}
 S&=&\int d^Dx \sqrt{-\tilde{G}}\bigg[\frac{\widetilde{R}}{2\kappa^2} - \frac{1}{2}\tilde{G}^{MN}\partial_{\mu}\xi \partial_{\nu}\xi 
 - \bigg(\frac{AF'(A) - F(A)}{2\kappa^2F'(A)^{D/(D-2)}}\bigg)\bigg]
 \label{transformation4}
\end{eqnarray}
Note that the field $\xi(x)$ acts as an scalar field with the potential $\frac{AF'(A) - F(A)}{2\kappa^2F'(A)^{D/(D-2)}}$ ($= V(A(\xi))$). Thus, the higher order curvature 
degree of freedom manifests itself as an scalar field degree of freedom $\xi(x)$ with the potential $V(\xi)$, which actually depends on the form of $F(R)$.

\section{Kalb-Ramond field in four dimension in $F(R)$ gravity}
\label{Model-I}

Let us firstly consider a four dimensional spacetime in $F(R)$ gravity. As mentioned earlier, here 
we are interested how the higher order terms affect the dynamical evolution of a second rank 
antisymmetric tensor field, generally known as Kalb-Ramond field ($B_{\mu\nu}$). Therefore the action of the model is given by, 

\begin{eqnarray}
 S&=&\int d^4x \sqrt{-g}\bigg[\frac{F(R)}{2\kappa^2} - \frac{1}{12}H_{\mu\nu\rho}H^{\mu\nu\rho}\bigg]\nonumber\\
 &=&\int d^4x \sqrt{-g}\bigg[\frac{1}{2\kappa^2}\bigg(R + \frac{R^2}{m^2}\bigg) - \frac{1}{12}H_{\mu\nu\rho}H^{\mu\nu\rho}\bigg]\ .
 \label{action1}
\end{eqnarray}

Here we are assuming a particular form for gravitational sector,  the so-called Starobinsky model \cite{Starobinsky:1980te}, $F(R)=R+\frac{R^2}{m^2}$, where $m$ is a parameter having mass dimension and 
$\frac{1}{2\kappa^2}=M_{(4)}^2$ ($M_{(4)}$ being the four dimensional 
Planck mass). Moreover, $H_{\mu\nu\alpha}$ is the field strength tensor of Kalb-Ramond (KR) field, defined by 
$H_{\mu\nu\alpha} = \partial_{[\mu}B_{\nu\alpha]}$. As we may notice that $H_{\mu\nu\alpha}$ is invariant 
under the KR gauge transformation: $B_{\mu\nu}\rightarrow B_{\mu\nu} + \partial_{[\mu}\omega_{\nu]}$ and thereby the action turns out also invariant under such transformation.\\
Using the conformal transformation from section \ref{sectionFR}, the action(\ref{action1}) can be expressed as an scalar-tensor theory:
\begin{eqnarray}
 S&=&\int d^4x \sqrt{-\tilde{g}}\bigg[\frac{\widetilde{R}}{2\kappa^2} - \frac{1}{2}\tilde{g}^{\mu\nu}\partial_{\mu}\xi \partial_{\nu}\xi 
 - V(\xi)\nonumber\\ 
 &-&\frac{1}{12}e^{-\sqrt{\frac{2}{3}}\kappa \xi} H_{\mu\nu\rho}H_{\alpha\beta\delta} 
 \tilde{g}^{\mu\alpha}\tilde{g}^{\nu\beta}\tilde{g}^{\rho\delta}\bigg]\ ,
 \label{action3}
\end{eqnarray}
where the scalar potential $V(\xi)$ has the following expression,
\begin{eqnarray}
 V(\xi) = \frac{m^2}{8\kappa^2}\bigg(1 - e^{\sqrt{\frac{2}{3}}\kappa\xi}\bigg)^2\ ,
 \label{potential}
\end{eqnarray}
The potential has an stable minima at $<\xi> = 0$ and asymptotically reaches $\frac{m^2}{8\kappa^2}$ as $\xi$ goes to $-\infty$. Fig.~\ref{plot_potential} depicts the form of the potential $V(\xi)$.
\begin{figure}[!h]
\begin{center}
 \centering
 \includegraphics[width=3.5in,height=2.0in]{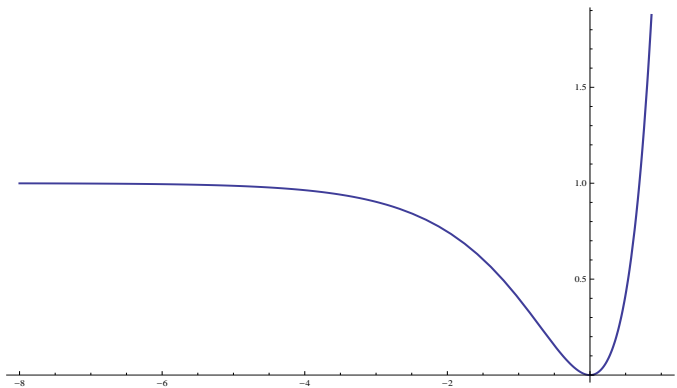}
 \caption{$V(\xi)$ vs $\xi$.}
 \label{plot_potential1}
\end{center}
\end{figure}
From eqn.(\ref{action3}), it is straightforward to show that the kinetic term of the KR field becomes non-canonical because of the presence of the scalar field $\xi(x)$. In order to make the 
KR field canonical, we redefine the field as follows :

\begin{eqnarray}
 B_{\mu\nu} \longrightarrow \widetilde{B}_{\mu\nu} = e^{-\frac{1}{2}\sqrt{\frac{2}{3}}\kappa \xi} B_{\mu\nu}\ .
 \label{redefine}
 \end{eqnarray}
 
Then, the final form of the scalar-tensor action can be expressed as follows:
 \begin{eqnarray}
 S = \int d^4x \sqrt{-\tilde{g}}\bigg[\frac{\widetilde{R}}{2\kappa^2} - \frac{1}{2}\tilde{g}^{\mu\nu}\partial_{\mu}\xi \partial_{\nu}\xi 
 - V(\xi) - \frac{1}{12} \tilde{H}_{\mu\nu\rho}\tilde{H}^{\mu\nu\rho}\bigg]\ ,
 \label{action4}
\end{eqnarray}
where we consider $\kappa\tilde{B}_{\mu\nu}<1$ and $V(\xi)$ is obtained from eqn.(\ref{potential}). In the following, 
we determine the solutions of the cosmological field equations for the scalar-tensor model (see eqn.(\ref{action4})), from which 
one can extract the corresponding solutions for the original $F(R)$ model (\ref{action1}) by taking the inverse conformal transformation.\\

\subsection{Cosmological field equations and solutions in the scalar-tensor representation}

In order to obtain the field equations of the scalar-tensor (ST) action (\ref{action4})), first we have to obtain the 
energy-momentum tensor for $\xi(x)$ and $\tilde{B}_{\mu\nu}(x)$,
\begin{eqnarray}
 T_{\mu\nu}[\xi]&=&\frac{2}{\sqrt{-\tilde{g}}}
 \frac{\delta}{\delta \tilde{g}^{\mu\nu}}\bigg[\sqrt{-\tilde{g}}\bigg(\frac{1}{2}\tilde{g}^{\alpha\beta}\partial_{\alpha}\xi\partial_{\beta}\xi + V(\xi)\bigg)\bigg]\nonumber\\
 &=&\partial_{\mu}\xi\partial_{\nu}\xi - \tilde{g}_{\mu\nu}\bigg(\frac{1}{2}\tilde{g}^{\alpha\beta}\partial_{\alpha}\xi\partial_{\beta}\xi + V(\xi)\bigg)\ ,
 \label{em tensor1}
\end{eqnarray}
and
\begin{eqnarray}
 T_{\mu\nu}[\tilde{B}]&=&\frac{2}{\sqrt{-\tilde{g}}}
 \frac{\delta}{\delta \tilde{g}^{\mu\nu}}\bigg[\frac{1}{12}\sqrt{-\tilde{g}}\tilde{g}^{\mu\alpha}\tilde{g}^{\nu\beta}\tilde{g}^{\lambda\gamma}
 \tilde{H}_{\mu\nu\lambda}\tilde{H}_{\alpha\beta\gamma}\bigg]\nonumber\\
 &=&\frac{1}{6}\bigg[3\tilde{g}_{\nu\rho}\tilde{H}_{\alpha\beta\mu}\tilde{H}^{\alpha\beta\rho} 
 - \frac{1}{2}\tilde{g}_{\mu\nu}\tilde{H}_{\alpha\beta\gamma}\tilde{H}^{\alpha\beta\gamma}\bigg]\ .
 \label{em tensor2}
\end{eqnarray}
Here we are interested in the cosmological evolution of the KR field. For that purpose, we assume the ansatz of a flat FRW metric:
\begin{eqnarray}
 \tilde{ds}^2&=&\tilde{g}_{\mu\nu}(x) dx^{\mu}dx^{\nu}\nonumber\\
 &=&-dt^2 + a^2(t)\big[dx^2 + dy^2 + dz^2\big]\ ,
 \label{4d metric}
\end{eqnarray}
where $t$ and $a(t)$ are the cosmic time and the scale factor respectively. However before obtaining the field equations, we would like 
to emphasize that $\tilde{H}_{\mu\nu\lambda}$ has 
four independent components in four dimensional spacetimes due to its antisymmetric nature, such that they can be expressed as,
\begin{eqnarray}
 \tilde{H}_{012} = h_1\ ,~~~~~~~~~~~~\tilde{H}^{012} = h^1\ ,\nonumber\\
 \tilde{H}_{013} = h_2\ ,~~~~~~~~~~~~\tilde{H}^{013} = h^2\ ,\nonumber\\
 \tilde{H}_{023} = h_3\ ,~~~~~~~~~~~~\tilde{H}^{023} = h^3\ ,\nonumber\\
 \tilde{H}_{123} = h_4\ ,~~~~~~~~~~~~\tilde{H}^{123} = h^4\ .
\label{independent}
\end{eqnarray}
As the KR field tensor $\tilde{H}_{\mu\nu\alpha}$ owns four independent components, can be 
equivalently expressed as a vector field (which has also four independent components in four dimensions) \cite{buchbinder}, 
$\tilde{H}_{\mu\nu\alpha} = \varepsilon_{\mu\nu\alpha\beta}\Upsilon^{\beta}$, with $\Upsilon^{\beta}$ being the vector field.\\
Moreover, eqn.~(\ref{independent}) together with the expressions for the energy-momentum tensor and the FLRW metric leads to the off-diagonal Friedmann equations as 
follows \cite{logR},
\begin{eqnarray}
 h_4h^3 = h_4h^2 = h_4h^1 = h_2h^3 = h_1h^3 = h_1h^2 = 0\ .
 \label{off einstein equation}
\end{eqnarray}
where the fields are considered homogeneous. The above set of equations has the following solution,
\begin{eqnarray}
 h_1 = h_2 = h_3 = 0~~~~~~~~~~~~~~~~~, h_4 \neq 0\ .
 \label{sol off einstein equation}
\end{eqnarray}
Using this solution, one easily obtains the total energy density and pressure for the matter fields ($\xi$, $\tilde{B}_{\mu\nu}$), 
which become $\rho_T = \bigg[\frac{1}{2}\dot{\xi}^2 + V(\xi) + \frac{1}{2}h_4h^4\bigg]$ and  $p_T = \bigg[\frac{1}{2}\dot{\xi}^2 - V(\xi) + \frac{1}{2}h_4h^4\bigg]$ respectively (where 
the dot denotes $\frac{d}{dt}$). As a result, the diagonal Friedmann equations turn out to be,
\begin{eqnarray}
 H^2 = \frac{\kappa^2}{3}\bigg[\frac{1}{2}\dot{\xi}^2 + \frac{m^2}{8\kappa^2}\big(1 - e^{\sqrt{\frac{2}{3}}\kappa\xi}\big)^2 
 + \frac{1}{2}h_4h^4\bigg]\ ,
 \label{einstein equation1}
\end{eqnarray}
and
\begin{eqnarray}
 2\dot{H} + 3H^2 + \kappa^2\bigg[\frac{1}{2}\dot{\xi}^2 
 - \frac{m^2}{8\kappa^2}\bigg(1 - e^{\sqrt{\frac{2}{3}}\kappa\xi}\bigg)^2 + \frac{1}{2}h_4h^4] = 0\ ,
 \label{einstein equation2}
\end{eqnarray}
where $H=\frac{\dot{a}}{a}$ is the Hubble parameter. In order to obtain the above equations, we have used the explicit expression of $V(\xi)$ as shown in eqn.~(\ref{potential}). 
Furthermore, the field equations for the KR field ($\tilde{B}_{\mu\nu}$) and the scalar field ($\xi$) are given by,
\begin{eqnarray}
 \tilde{\nabla}_{\mu}\tilde{H}^{\mu\nu\lambda} = \frac{1}{a^3(t)}\partial_{\mu}\bigg[a^3(t)\tilde{H}^{\mu\nu\lambda}\bigg] = 0\ ,
 \label{KR equation}
\end{eqnarray}
and
\begin{eqnarray}
 \ddot{\xi} + 3H\dot{\xi} - \sqrt{\frac{2}{3}}\frac{m^2}{4\kappa}e^{\sqrt{\frac{2}{3}}\kappa\xi}\big(1 - e^{\sqrt{\frac{2}{3}}\kappa\xi}\big)  = 0\ .
 \label{scalar equation}
\end{eqnarray}
From eqn.~(\ref{KR equation}), we know that the non-zero component of $\tilde{H}_{\mu\nu\alpha}$ 
(i.e $\tilde{H}_{123}=h_4$) depends on $t$ only (see Appendix-I for the derivation), which is also expected from the gravitational field equations. 
Differentiating both sides (with respect to $t$) of eqn.~(\ref{einstein equation1}), one easily obtains
\begin{eqnarray}
6H\dot{H} = \kappa^2\bigg[\dot{\xi}\ddot{\xi} 
- \sqrt{\frac{2}{3}}\frac{m^2}{4\kappa}e^{\sqrt{\frac{2}{3}}\kappa\xi}\big(1 - e^{\sqrt{\frac{2}{3}}\kappa\xi}\big)\dot{\xi} 
+ \frac{1}{2}\frac{d}{dt}(h_4h^4)\bigg]
\nonumber
\end{eqnarray}

Furthermore, eqns.~(\ref{einstein equation1}) and (\ref{einstein equation2}) lead to the expression $2\dot{H} = -\kappa^2[\dot{\xi}^2 + h_4h^4]$. Substituting this expression of $\dot{H}$ 
in the above equation and using the scalar field equation of motion, we obtain the following cosmic evolution for $h_4h^4$:
\begin{eqnarray}
 \frac{d}{dt}(h_4h^4) = -6H h_4h^4\ .
 \label{evolution of KR energy density}
\end{eqnarray}
Solving (\ref{evolution of KR energy density}), we get
\begin{eqnarray}
 h_4h^4 = \frac{h_0}{a^6}\ ,
 \label{solution of KR energy density}
\end{eqnarray}
with $h_0$ being an integration constant which must be taken positive in order to ensure a real solution for $h^4(t)$. Recall that 
the term $\frac{1}{2}h_4h^4$ represents the energy density contribution from the Kalb-Ramon field. Therefore, eqn.~(\ref{solution of KR energy density}) 
clearly indicates that the KR field energy density ($\tilde{\rho}_{KR}$) is proportional to $1/a^6$ and as a result, $\tilde{\rho}_{KR}$ gradually 
decreases with the expansion of the universe. However, eqn.~(\ref{solution of KR energy density}) also shows that KR energy density is large 
and may play a significant role in the early universe (when the scale factor is small in comparison to the present one). Therefore, 
in order to address the dynamical suppression of KR field, we should study the dynamics of KR field from the very early universe when 
it is also important to evaluate whether the early universe undergoes through an accelerating stage or not, i.e. an inflationary phase. To check this 
phenomena, we need to obtain the form of the scale factor at early times.\\
From the solution $h_4h^4$ in terms of the scale factor, two independent equations remained,
\begin{eqnarray}
 H^2 = \frac{\kappa^2}{3}\bigg[\frac{1}{2}\dot{\xi}^2 + \frac{m^2}{8\kappa^2}\big(1 - e^{\sqrt{\frac{2}{3}}\kappa\xi}\big)^2 
 + \frac{1}{2}\frac{h_0}{a^6}\bigg]\ ,
 \label{independent equation1}
\end{eqnarray}
and
\begin{eqnarray}
 \ddot{\xi} + 3H\dot{\xi} - \sqrt{\frac{2}{3}}\frac{m^2}{4\kappa}e^{\sqrt{\frac{2}{3}}\kappa\xi}\big(1 - e^{\sqrt{\frac{2}{3}}\kappa\xi}\big) = 0\ .
 \label{independent equation2}
 \end{eqnarray}
Here, we should mention that eqns.~(\ref{independent equation1}) and (\ref{independent equation2}) match to the field equations when 
$\tilde{H}_{\mu\nu\alpha}$ is expressed in the vector representation i.e 
$\tilde{H}_{\mu\nu\alpha}=\varepsilon_{\mu\nu\alpha\beta}\Upsilon^{\beta}$ (see Appendix \ref{appendix-II} 
for the derivation of this equivalence). This confirms the equivalence between the two representations 
at the level of the equations of motion, which is also in agreement with Ref.~\cite{buchbinder}.\\
However, eqns.~(\ref{independent equation1}) and (\ref{independent equation2}) are sufficient to determine the evolution of two unknowns functions: the scale factor $a(t)$ and the scalar 
field $\xi(t)$. As mentioned above, we are interested in solving the field equations during the initial phase of the universe where the potential energy of the scalar field 
is assumed to be greater than that of the kinetic energy, known as the slow-roll approximation i.e.
 \begin{eqnarray}
 V(\xi) \gg \frac{1}{2}\dot{\xi}^2\ .
 \label{slow roll approximation}
\end{eqnarray}
Under such approximation, eqns.~(\ref{independent equation1}) and (\ref{independent equation2}) become,
\begin{eqnarray}
 H^2 = \frac{\kappa^2}{3}\bigg[\frac{m^2}{8\kappa^2}\big(1 - e^{\sqrt{\frac{2}{3}}\kappa\xi}\big)^2 + \frac{1}{2}\frac{h_0}{a^6}\bigg]\ ,
 \label{slow roll equation1a}
\end{eqnarray}
and
\begin{eqnarray}
 3H\dot{\xi} - \sqrt{\frac{2}{3}}\frac{m^2}{4\kappa}e^{\sqrt{\frac{2}{3}}\kappa\xi}\big(1 - e^{\sqrt{\frac{2}{3}}\kappa\xi}\big) = 0\ .
 \label{slow roll equation2a}
\end{eqnarray}
By considering $\frac{\kappa^2h_0}{m^2}<1$ (which is also necessary in order to relate the model with the observational constraints, as 
shown below), eqns.~(\ref{slow roll equation1a}) and (\ref{slow roll equation2a}) can be written as follows,
\begin{eqnarray}
 H = \frac{m}{2\sqrt{6}}\big(1 - e^{\sqrt{\frac{2}{3}}\kappa\xi}\big) 
 \bigg[1 + \frac{2\kappa^2h_0}{m^2} \frac{1}{a^6\big(1 - e^{\sqrt{\frac{2}{3}}\kappa\xi}\big)^2}\bigg]\ ,
 \label{slow roll equation1b}
\end{eqnarray}
and
\begin{eqnarray}
 \frac{d\xi}{dt} = \frac{m}{18\kappa}e^{\sqrt{\frac{2}{3}}\kappa\xi} 
 \bigg[1 - \frac{2\kappa^2h_0}{m^2} \frac{1}{a^6\big(1 - e^{\sqrt{\frac{2}{3}}\kappa\xi}\big)^2}\bigg]\ ,
 \label{slow roll equation2b}
\end{eqnarray}
where we keep the terms up to the leading order in $h_0$. Under the condition $\frac{\kappa^2h_0}{m^2} < 1$, we can solve the above equations 
for $\xi(t)$ and $a(t)$) perturbatively where $\frac{\kappa^2h_0}{m^2}$ is treated as a perturbation parameter. The solutions are (for 
$m_0\neq0$):
\begin{eqnarray}
 \xi(t) = \sqrt{\frac{3}{2\kappa^2}}\bigg[\ln{\bigg(\frac{9}{-\sqrt{6}m(t-t_0) + 9C}\bigg)} + \frac{\kappa^2h_0}{m^2}P(t)\bigg]\ ,
 \label{solution scalar field ST}
\end{eqnarray}
and
\begin{eqnarray}
 a(t) = D\bigg(1 - \frac{2m(t-t_0)}{3\sqrt{6}C}\bigg)^{3/4} 
 \bigg(1 + \frac{\kappa^2h_0}{m^2}Q(t)\bigg) \exp{\bigg[\frac{m(t-t_0)}{2\sqrt{6}}\bigg]}\ .
 \label{solution scale factor ST}
\end{eqnarray}
Here $P(t)$ and $Q(t)$ have the following expressions,
\begin{eqnarray}
 P(t) = \frac{(-\sqrt{6}m(t-t_0) + 9C)^2}{(-\sqrt{6}m(t-t_0) + 9C + 9)^3}\ ,
 \label{U}
\end{eqnarray}
and
\begin{eqnarray}
 Q(t)=
 \frac{\big(5 + 9C(1+3C)\big)\big(1 - \sqrt{\frac{3}{2}}m(t-t_0)\big) + \sqrt{6}\big(1 + 6C + 81C^2\big)
 \big(-m(t-t_0) + \sqrt{\frac{3}{2}}m^2(t-t_0)^2\big)}{\bigg(1 - \frac{2m(t-t_0)}{3\sqrt{6}C}\bigg)^{7/2}}\ .
 \label{V}
\end{eqnarray}
Furthermore, $C$ and $D$ are integration constants related to the initial values of $\xi(t)$ and $a(t)$ as follows :
\begin{eqnarray}
 \xi(t_0)&=&\xi_0=\sqrt{\frac{3}{2\kappa^2}} \bigg[\ln{(1/C)} + \frac{\kappa^2h_0}{9m^2}\frac{C^2}{(1+C)^3}\bigg]\ ,\nonumber \\
 a(t_0) &=& D \bigg[1 + \frac{\kappa^2h_0}{m^2}\big(5 + 9C(1+3C)\big)\bigg]\ .
 \label{initial value of scale factor ST}
\end{eqnarray}
Note that for $h_0\rightarrow 0$, both solutions $\xi(t)$ as $a(t)$ go towards the well known Starobinsky solution. Furthermore, 
eqn.~(\ref{solution scalar field ST}) leads to the fact that the scalar field increases with time and goes to infinity as 
$t\rightarrow \frac{9C}{\sqrt{6}m}$. Keeping this in mind, here we consider the initial value of the scalar field (i.e $\xi_0$) as negative. 
The negative initial value of the scalar field is also consistent with the slow roll condition, as may be noticed from Fig.~\ref{plot_potential1}.

\subsubsection{Beginning of inflation in scalar-tensor model}

After obtaining the solution of the scale factor (\ref{solution scale factor ST}), we can now analyse whether this form of the scale factor corresponds to 
an accelerating stage during the early universe (i.e $t\gtrsim t_0$). For this purpose, we expand $a(t)$ in the form of Taylor series 
(about $t=t_0$) and keep those terms up to linear order in $t-t_0$:

\begin{eqnarray}
 a(t \rightarrow t_0) = D\big(1 - \frac{2m(t-t_0)}{3\sqrt{6}C}\big)^{3/4} \exp{\big[\frac{m(t-t_0)}{2\sqrt{6}}\big]}\nonumber\\
 \bigg[1 + \frac{\kappa^2h_0}{m^2}\bigg(\big(5 + 9C(1+3C)\big) - \frac{m}{\sqrt{6}}\big(11 + 45C + 513C^2\big)t\bigg)\bigg]\ ,
 \label{limiting scale factor ST}
\end{eqnarray}
where we have used the expression of $Q(t)$ at $t\rightarrow t_0$ as $Q(t \rightarrow t_0) = \bigg(5 + 9C(1+3C)\bigg) - \frac{m}{\sqrt{6}}\bigg(11 + 45C + 513C^2\bigg)t$. Differentiating (twice) 
both sides of 
eqn.~(\ref{limiting scale factor ST}) in the limit $t\rightarrow t_0$, one finally get the following expression for the acceleration:
\begin{eqnarray}
 \frac{\ddot{a}}{a}\bigg|_{t\rightarrow t_0} = \big(\frac{m}{2\sqrt{6}}\big)^2 \frac{C-1}{C} \bigg[\frac{C-1}{C} 
 - 4\frac{\kappa^2h_0}{m^2}\big(11 + 45C + 513C^2\big)\bigg]\ .
 \label{limiting acceleration ST}
\end{eqnarray}
By inverting eqn.~(\ref{initial value of scale factor ST}), we obtain the explicit expression for the integration constant $C$ in terms 
of $\xi_0$ (initial value of the scalar field). For the zeroth order in $h_0$, one gets $C^{(0)} = e^{-\sqrt{\frac{2\kappa^2}{3}}\xi_0}$ and up to first order in $h_0$, $C$ becomes,
\begin{eqnarray}
 C&=&e^{-\sqrt{\frac{2\kappa^2}{3}}\xi_0} + \frac{\kappa^2h_0}{9m^2}\bigg(\frac{e^{-3\sqrt{\frac{2\kappa^2}{3}}\xi_0}}{(1 + e^{-\sqrt{\frac{2\kappa^2}{3}}\xi_0})^3}\bigg)\nonumber\\
 &=&e^{|\sigma_0|} + \frac{\kappa^2h_0}{9m^2}\bigg(\frac{e^{3|\sigma_0|}}{(1 + e^{|\sigma_0|})^3}\bigg)\ ,
 \label{C}
\end{eqnarray}
where $\sigma_0=\sqrt{\frac{2\kappa^2}{3}\xi_0}$ and recall that the initial value of the scalar field is considered to be negative. Then, by using the 
above expression for $C$ (see eqn.(\ref{C})), eqn.~(\ref{limiting acceleration ST}) turns out:
\begin{eqnarray}
 \frac{\ddot{a}}{a}\bigg|_{t\rightarrow t_0} = \big(\frac{m}{2\sqrt{6}}\big)^2 (1-e^{-|\sigma_0|}) \bigg[\big(1-e^{-|\sigma_0|}\big) 
 - 4\frac{\kappa^2h_0}{m^2}\big(11 + 45e^{|\sigma_0|} + 513e^{2|\sigma_0|} - \frac{e^{|\sigma_0|}}{18(1 + e^{|\sigma_0|})^3}\big)\bigg]\ .
 \label{limiting acceleration ST1}
\end{eqnarray}
Note that under the condition
\begin{eqnarray}
 \frac{m^2(1 - e^{-|\sigma_0|})}{4\big(11 + 45e^{|\sigma_0|} + 513e^{2|\sigma_0|} - \frac{e^{|\sigma_0|}}{18(1 + e^{|\sigma_0|})^3}\big)} > \kappa^2h_0\ ,
 \label{condition ST}
\end{eqnarray}
the universe passes through an acceleration phase while it does not when the 
condition $\frac{m^2(1 - e^{-|\sigma_0|})}{4\big(11 + 45e^{|\sigma_0|} + 513e^{2|\sigma_0|} - \frac{e^{|\sigma_0|}}{18(1 + e^{|\sigma_0|})^3}\big)} < \kappa^2h_0$ holds.\\

Hence, the parameters $m$ and $h_0$ control the strength of the scalar field and the KR field energy density respectively. Therefore, the interplay among the scalar field and the 
KR field fixes whether the early universe evolves through an accelerating stage or not. In the next section, we focus again on the cosmological solutions and their possible 
consequences for the original $F(R)$ model (see eqn.(\ref{action1})) by using the solutions of the corresponding scalar-tensor theory.

\subsection{Cosmological solutions and their possible consequences in the $F(R)$ gravity: Suppression of the Kalb-Ramond field}

Recall that the original higher order curvature $F(R)$ model is 
given by the action (\ref{action1}), solutions for the metric can be obtained from the corresponding scalar-tensor theory (see eqns. (\ref{solution scalar field ST}) 
and (\ref{solution scale factor ST})) with the help of the inverse conformal transformation. Thus, the line element in $F(R)$ model can be written as:
\begin{eqnarray}
 ds^2&=&e^{\sqrt{\frac{2}{3}}\kappa\xi(t)} \bigg[-dt^2 + a^2(t)\big(dx^2 + dy^2 + dz^2\big)\bigg]\nonumber\\
 &=&-d\tau^2 + s^2(\tau)\big(dx^2 + dy^2 + dz^2\big)\ , 
\end{eqnarray}
where $\tau(t)$, $s(\tau)$ are the cosmic time and scale factor respectively in Jordan frame, which are related to the Einstein frame by the conformal transformation:
\begin{eqnarray}
\tau(t) = \int dt e^{[\frac{1}{2}\sqrt{\frac{2}{3}}\kappa\xi(t)]}\ ,
\label{cosmic time1 F(R)}
\end{eqnarray}
and
\begin{eqnarray}
 s(\tau(t)) = e^{[\frac{1}{2}\sqrt{\frac{2}{3}}\kappa\xi(t)]} a(t)\ .
 \label{scale factor1 F(R)}
\end{eqnarray}
Eqn.(\ref{cosmic time1 F(R)}) clearly indicates that $\tau(t)$ is a monotonically increasing function of $t$. However, by integrating eqn.~(\ref{cosmic time1 F(R)}), one gets 
the explicit functional form of $\tau(t)$ as follows,
\begin{eqnarray}
 \tau - \tau_0&=&\frac{1}{4m}\sqrt{\frac{3}{2}} \bigg[\bigg(8\sqrt{9C} - 8\sqrt{9C - \sqrt{6}m(t-t_0)}\bigg)\ ,\nonumber\\
 &+&\frac{\kappa^2h_0}{2m^2}\bigg(\frac{(27 + 45C - 5\sqrt{6}m(t-t_0))\sqrt{9C - \sqrt{6}m(t-t_0)}}{(9 + 9C - \sqrt{6}m(t-t_0))^2} 
 - \frac{(3+5C)\sqrt{C}}{3(1+C)^2}\bigg)\ ,\nonumber\\
 &+&\frac{\kappa^2h_0}{2m^2}\bigg(\tan^{-1}\big(\frac{3}{\sqrt{9C - \sqrt{6}m(t-t_0)}}\big) - \tan^{-1}(1/\sqrt{C})\bigg)\bigg]\ ,
 \label{cosmic time2 F(R)}
\end{eqnarray}
where the integration constant (that appeared while integrating eqn.~(\ref{cosmic time1 F(R)})) is fixed by the condition $\tau(t_0) = \tau_0$. 
Moreover, with the solutions of $\xi(t)$ and $a(t)$, eqn.(\ref{scale factor1 F(R)}) immediately leads to the form of $s(\tau)$ (in terms of $t$, where 
$\tau(t)$ is given by the above expression) as,
\begin{eqnarray}
 s(\tau(t))&=&D\bigg(1 - \frac{2m(t-t_0)}{3\sqrt{6}C}\bigg)^{3/4} 
 \bigg(1 + \frac{\kappa^2h_0}{m^2}\big(Q(t) + \frac{1}{2}P(t)\big)\bigg)\ ,\nonumber\\ 
 &\exp&{\bigg[\frac{m(t-t_0)}{2\sqrt{6}} + \ln{\bigg(\frac{9}{-\sqrt{6}m(t-t_0) + 9C}\bigg)}\bigg]}\ ,
 \label{scale factor2 F(R)}
\end{eqnarray}
where $P(t)$ and $Q(t)$ are given by eqns.(\ref{U}) and (\ref{V}) respectively. However, by using eqn.~(\ref{cosmic time2 F(R)}), we obtain 
$\tau(t)$ at $t\rightarrow t_0$,
\begin{eqnarray}
 \tau(t\rightarrow t_0) = \tau_0 + \sqrt{\frac{1}{C}}\bigg[1 + \frac{\kappa^2h_0}{9m^2}\big(\frac{C^2}{(1+C)^3}\big)\bigg](t - t_0)\ ,
 \label{limiting cosmic time F(R)}
\end{eqnarray}

\subsubsection{Beginning of inflation in $F(R)$ gravity}

In this section, we investigate whether the solution of the scale factor ($s(\tau)$, see eqn.(\ref{scale factor2 F(R)})) corresponds 
to an inflationary stage of the early universe. In order to analyse this matter, we expand $s(\tau)$ in the form of a
Taylor series (about $\tau=\tau_0$) and keep the terms up to linear order in $\tau-\tau_0$. For this purpose, we need the the expression of 
$\tau(t)$ at $t\rightarrow t_0$, which can be obtained from eqn.~(\ref{cosmic time2 F(R)}) as,

\begin{eqnarray}
 \tau(t\rightarrow t_0) = \tau_0 + \sqrt{\frac{1}{C}}\bigg[1 + \frac{\kappa^2h_0}{9m^2}\big(\frac{C^2}{(1+C)^3}\big)\bigg](t - t_0)\ .
 \label{limiting cosmic time F(R)}
\end{eqnarray}
Recall that $\tau$ goes to $\tau_0$ as $t\rightarrow t_0$, which is also evident from the above expression. Eqns.(\ref{scale factor2 F(R)}) and (\ref{limiting cosmic time F(R)}) lead 
to the expression of the scale factor at $\tau\rightarrow \tau_0$,
\begin{eqnarray}
 s(\tau \rightarrow \tau_0)&=&\frac{D}{\sqrt{C}}\bigg[1 - \frac{2m\beta(\tau-\tau_0)}{3\sqrt{6}C}\bigg]^{3/4} 
 \exp{\bigg[\frac{m\beta(\tau-\tau_0)}{2\sqrt{6}}\bigg]} \bigg[1 + \frac{\kappa^2h_0}{m^2}\bigg(\big[5 + 9C(1+3C)\big]\ ,\nonumber\\ 
 &+&\frac{C^2}{9(1+C)^3} - \frac{m\beta}{\sqrt{6}}\big[11 + \frac{1219}{27}C + \frac{13849}{27}C^2\big](\tau-\tau_0)\bigg)\bigg]\ ,
 \label{limiting scale factor F(R)}
\end{eqnarray}
with $\beta$ given by,
\begin{eqnarray}
 1/\beta = \sqrt{\frac{1}{C}}\bigg[1 + \frac{\kappa^2h_0}{9m^2}\big(\frac{C^2}{(1+C)^3}\big)\ ,
 \nonumber
\end{eqnarray}
Differentiating twice both sides with respect to $\tau$, eqn.~(\ref{limiting scale factor F(R)}) becomes:
\begin{eqnarray}
 \frac{1}{s}\frac{d^2s}{d\tau^2}\bigg|_{\tau\rightarrow \tau_0} = \beta^2 \big(\frac{m}{2\sqrt{6}}\big)^2 (1-e^{-|\sigma_0|}) \bigg[\big(1-e^{-|\sigma_0|}\big)
 - 4\frac{\kappa^2h_0}{m^2}\big(11 + \frac{1219}{27}e^{|\sigma_0|} + \frac{13849}{27}e^{2|\sigma_0|}\big)\bigg]\ ,
 \label{limiting acceleration F(R)}
\end{eqnarray}
where $C= e^{|\sigma_0|} + \frac{\kappa^2h_0}{9m^2}\bigg(\frac{e^{3|\sigma_0|}}{(1 + e^{|\sigma_0|})^3}\bigg)$ is used. Therefore, it is clear that for,
\begin{eqnarray}
 \frac{m^2(1 - e^{-|\sigma_0|})}{4\big(11 + \frac{1219}{27}e^{|\sigma_0|} + \frac{13849}{27}e^{2|\sigma_0|} - \frac{e^{|\sigma_0|}}{18(1 + e^{|\sigma_0|})^3}\big)} > \kappa^2h_0\ ,
 \label{condition F(R)}
\end{eqnarray}
the early universe undergoes through an inflationary stage (with $\tau_0$ as the onset of inflation), while for  
$\frac{m^2(1 - e^{-|\sigma_0|})}{4\big(11 + \frac{1219}{27}e^{|\sigma_0|} + \frac{13849}{27}e^{2|\sigma_0|} - \frac{e^{|\sigma_0|}}{18(1 + e^{|\sigma_0|})^3}\big)} < \kappa^2h_0$, 
$\frac{d^2s}{d\tau^2}\bigg|_{\tau\rightarrow \tau_0}$ becomes negative.\\ 
Comparison of eqns.~(\ref{condition ST}) and (\ref{condition F(R)}) makes it clear that the conditions for an early time acceleration 
in scalar-tensor theory and $F(R)$ gravity are different. However, note that similarly as in scalar-tensor theory, the interplay among the parameters $m$ and $h_0$ fixes whether the 
universe evolves through an inflationary stage during the early universe. Furthermore, in order to solve the flatness and horizon problems, the universe must passes through an 
accelerating stage at early epoch (in the original $F(R)$ model) and from this requirement, here we assume the condition shown in eqn.~(\ref{condition F(R)}).

\subsubsection{End of inflation in $F(R)$ gravity}

In the previous section, we have shown that the very early universe expands with acceleration, a phase generally known as the inflationary epoch. At this stage, it is important to 
check whether the inflationary era has an end in a finite time. We may define the end of inflation by the condition:
\begin{eqnarray}
 \frac{d^2s}{d\tau^2} = 0\ .
 \label{end0}
\end{eqnarray}
Recall $s(\tau) = e^{[\sqrt{\frac{1}{6}\kappa\xi(t)}]}a(t)$ is the scale factor in the $F(R)$ model, from which one obtains,
\begin{eqnarray}
 \frac{d^2s}{d\tau^2} = e^{[-\frac{1}{2}\sqrt{\frac{2}{3}\kappa\xi(t)}]} a(t) \bigg[\dot{H} + H^2 + 
 \frac{\kappa}{\sqrt{6}}H\dot{\xi}\bigg]\ ,
 \label{end1}
\end{eqnarray}
where $H$ is the Hubble parameter in the scalar-tensor picture and the dot represents $\frac{d}{dt}$. Eqns.~(\ref{end0}) and (\ref{end1}) clearly indicate that the end of the 
inflationary epoch in the $F(R)$ model can be expressed by the following equation,
\begin{eqnarray}
 \dot{H} + H^2 + \frac{\kappa}{\sqrt{6}}H\dot{\xi} = 0\ .
 \label{end2}
\end{eqnarray}
Now we analyse whether this condition is consistent with the field equations. Differentiating (with respect to $t$) both sides of 
eqn.~(\ref{slow roll equation1a}), we get,
\begin{eqnarray}
 \dot{H} = -\frac{m^2}{18}e^{2\sqrt{\frac{2}{3}}\kappa\xi(t)} + \frac{1}{12H} \frac{d}{dt}\big(\frac{\kappa^2h_0}{a^6(t)}\big)\ .
 \label{end3}
\end{eqnarray}
 Here we have used the scalar field equation. At the end of inflation, the term proportional to $\frac{\kappa^2h_0}{a^6}$ becomes small enough so that we can apply the method of 
 iteration (with respect to that term) to determine $\dot{H}$. Up to zeroth order of iteration,  $\dot{H}=-\frac{m^2}{18}e^{[2\sqrt{\frac{2}{3}}\kappa\xi(t)]}$. Consequently, one 
 determines $\dot{H}$ up to first order of iteration as 
 follows:
 \begin{eqnarray}
  \dot{H} = -\frac{m^2}{18}e^{2\sqrt{\frac{2}{3}}\kappa\xi(t)} 
  + \big(\frac{2m^2}{81}\big)\frac{e^{4\sqrt{\frac{2}{3}}\kappa\xi(t)}}{\big(1 - \sqrt{\frac{2}{3}}\kappa\xi(t)\big)^2}\ .
 \label{end4}
 \end{eqnarray}
 By this expression together with the field equations of motion, eqn.~(\ref{end4}) leads to the following condition on 
 the scalar field,
  \begin{eqnarray}
  \frac{1}{4} - \frac{2}{3}e^{[\sqrt{\frac{2}{3}}\kappa\xi(t_f)]} + \frac{1}{9}e^{[2\sqrt{\frac{2}{3}}\kappa\xi(t_f)]} 
  + \frac{8}{81}e^{[4\sqrt{\frac{2}{3}}\kappa\xi(t_f)]} = 0\ ,
  \nonumber
 \end{eqnarray}
 where $t_f-t_0$ denotes the duration of inflation in the $F(R)$ model (in terms of $t$). Solving the above algebraic equation (for $\xi(t_f)$), 
 we obtain 
 \begin{eqnarray}
  \xi_{f} \simeq \sqrt{\frac{3}{2\kappa^2}} \ln{\bigg(\frac{3}{5}\bigg)}\ .
  \label{final value of scalar field}
 \end{eqnarray}
 Eqn.~(\ref{final value of scalar field}) clearly indicates that the inflationary era of the universe continues as long 
 as the value of the scalar field remains greater than $\xi_f$ ($= \sqrt{\frac{3}{2\kappa^2}} \ln{\bigg(\frac{3}{5}\bigg)} < 0$). 
 Correspondingly the duration of inflation  can be calculated from the solution of the scalar field (see eqn.~\ref{solution scalar field ST})) as follows :
  \begin{eqnarray}
  t_f - t_0 = \frac{9}{\sqrt{6}m} \bigg[e^{|\sigma_0|} - e^{|\sigma_f|} \bigg(1 + \frac{\kappa^2h_0}{9m^2}e^{-|\sigma_0|}\bigg)\bigg]\ ,
  \label{duration1}
 \end{eqnarray}
  where $\sigma_0 = \sqrt{\frac{2\kappa^2}{3}}\xi_0$ and $\sigma_f = \sqrt{\frac{2\kappa^2}{3}}\xi_f$. Therefore in terms 
 of the cosmic time $\tau$, the duration of inflation becomes 
 \begin{eqnarray}
  \tau_f - \tau_0&=&\frac{1}{4m}\sqrt{\frac{3}{2}} \bigg[\bigg(8\sqrt{9C} - 8\sqrt{d}\bigg) 
  + \frac{\kappa^2h_0}{2m^2}\bigg(\frac{(27 + 5d)\sqrt{d}}{(9 + d)^2} 
 - \frac{(3+5C)\sqrt{C}}{3(1+C)^2}\bigg)\ ,\nonumber\\
 &+&\frac{\kappa^2h_0}{2m^2}\bigg(\tan^{-1}\big(\frac{3}{\sqrt{q}}\big) - \tan^{-1}(1/\sqrt{C})\bigg)\bigg]\ ,
 \label{duration2}
 \end{eqnarray}
  with $d = e^{\sigma_f}\bigg(1 + \frac{\kappa^2h_0}{9m^2}e^{-|\sigma_0|}\bigg)$. Moreover, in order to derive the above expression, we have used 
 eqn.~(\ref{cosmic time2 F(R)}). Note that $\tau_f-\tau_0$ depends on the parameters $\frac{\kappa^2h_0}{m^2}$ and $\sigma_0$. Therefore, we need the values of such parameters to 
 estimate the duration explicitly, which can be determined from the expression 
 of the spectral index and the tensor to scalar ratio as discussed in the next section.
 
 \subsubsection{Spectral Index, tensor to scalar ratio and number of e-foldings in $F(R)$ model}
%
 In order to test the broad inflationary paradigm as well as particular models versus the observations, we need 
 to calculate the value of spectral index($n_s$) and tensor to scalar ratio ($r$) and for this purpose, here we define a 
 dimensionless parameter (known as slow roll parameter) as,
 \begin{eqnarray}
  \epsilon_{F} = -\frac{1}{H_F^2}\frac{dH_F}{d\tau}\ ,
  \label{slow roll parameter F(R)1}
 \end{eqnarray}
where $H_F$ is the Hubble parameter in the $F(R)$ model, defined as $H_{F}=\frac{1}{s(\tau)}\frac{ds}{d\tau}$. By eqns.~(\ref{cosmic time1 F(R)}) 
 and (\ref{scale factor1 F(R)}), $H_F$ turns out to be,
 \begin{eqnarray}
  H_{F} = e^{-\frac{\sqrt{2}\kappa}{2\sqrt{3}}\xi(t)}\bigg[H + \frac{\sqrt{2}\kappa}{2\sqrt{3}}\dot{\xi}\bigg]\ .
  \label{hubble F(R)}
 \end{eqnarray}
 Recall that $H$ is the Hubble parameter in the corresponding scalar-tensor frame. In order to find the explicit expression of $\epsilon_F$, 
 we determine $\frac{dH_F}{d\tau}$ by differentiating with respect to $\tau$, both sides of eqn.~(\ref{hubble F(R)}), leading to:
 \begin{eqnarray}
  \frac{dH_{F}}{d\tau} = e^{-\frac{\sqrt{2}\kappa}{\sqrt{3}}\xi(t)}\bigg[\dot{H} - \frac{\sqrt{2}\kappa}{2\sqrt{3}}H\dot{\xi}\bigg]\ .
  \label{hubble derivative F(R)}
 \end{eqnarray}
  From the above expressions of $H_F$ and $\frac{dH_F}{d\tau}$ along with the slow roll field equations, one finally get the following 
 form of $\epsilon_F$,
 \begin{eqnarray}
  \epsilon_{F}&=&-\frac{1}{H_F^2}\frac{dH_F}{d\tau}\ ,\nonumber\\
  &=&\bigg[\frac{\frac{3\kappa^2h_0}{2a^6} + \frac{m^2}{12}e^{\sigma}(1 - e^{\sigma})}
  {\frac{m^2}{8}(1 - e^{\sigma})^2 + \frac{\kappa^2h_0}{2a^6} + \frac{m^2}{6}e^{\sigma}(1 - e^{\sigma})}\bigg]\ ,
  \label{slow roll parameter F(R)2}
 \end{eqnarray}
 with $\sigma(\tau) = \sqrt{\frac{2\kappa^2}{3}}\xi\big(t(\tau)\big)$.\\ 
 As mentioned above, the second rank antisymmetric KR field can be equivalently expressed as a vector field which can be further 
 recast as the derivative of a massless scalar field (see Appendix-II). As a consequence, 
 the spectral index and tensor to scalar ratio in the present context are defined as follows \cite{1,2,3}:
 \begin{eqnarray}
  n_s = \big[1 - 4\epsilon_F - 2\epsilon_2 + 2\epsilon_3 - 2\epsilon_4\big]\bigg|_{\tau_0}\ ,
  \label{spectral index1}
 \end{eqnarray}
 and
 \begin{eqnarray}
  r = 8\kappa^2 \frac{\varTheta}{F'(R)}\bigg|_{\tau_0}\ .
  \label{ratio1}
 \end{eqnarray}
Here the slow roll parameters ($\epsilon_F$, $\epsilon_2$, $\epsilon_3$, $\epsilon_4$) are defined by the following expressions,
\begin{eqnarray}
 \epsilon_F&=&-\frac{1}{H_F^2} \frac{dH_F}{d\tau}\ ,~~~~~~~~~~\epsilon_2 = \frac{1}{2\rho_{KR}H_F} \frac{d\rho_{KR}}{d\tau}\ ,\nonumber\\
 \epsilon_3&=&\frac{1}{2F'(R)H_F} \frac{dF'(R)}{d\tau}\ ,~~~~~~~~~~\epsilon_4 = \frac{1}{2EH_F} \frac{dE}{d\tau}\ ,
 \label{various slow roll parameters}
\end{eqnarray}
where $\varTheta$ and $E$ are given by,
\begin{eqnarray}
 \varTheta = \frac{\rho_{KR}}{F'(R)H_F^2}\bigg[F'(R) + \frac{3}{2\kappa^2\rho_{KR}}\bigg(\frac{d}{d\tau}F'(R)\bigg)^2\bigg]\ ,
 \label{vartheta}
\end{eqnarray}
and
\begin{eqnarray}
 E = \frac{\varTheta F'(R)H_F^2}{\rho_{KR}}\ ,
 \label{E}
\end{eqnarray}
with $\rho_{KR}$ ($= H_{123}H^{123}$) being the energy density of the KR field in the $F(R)$ model. However, by virtue of 
eqn.~(\ref{energy density1}), the variation of $\rho_{KR}$ immediately lead to 
$\rho_{KR} = e^{-2\sqrt{\frac{2}{3}}\kappa\xi(t)} \frac{h_0}{a^6}$. Keeping this in mind, now we are going to determine the explicit expressions of 
various terms appearing in the right hand side of eqns.~(\ref{spectral index1}) and (\ref{ratio1}). 
\begin{itemize}
 \item \underline{$\epsilon_F$}: As obtained in eqn.~(\ref{slow roll parameter F(R)2}), $\epsilon_F$ is given by
  \begin{eqnarray}
  \epsilon_{F} = \bigg[\frac{\frac{3\kappa^2h_0}{2a^6} + \frac{m^2}{12}e^{\sigma}(1 - e^{\sigma})}
  {\frac{m^2}{8}(1 - e^{\sigma})^2 + \frac{\kappa^2h_0}{2a^6} + \frac{m^2}{6}e^{\sigma}(1 - e^{\sigma})}\bigg]\ .
  \label{first slow roll parameter}
 \end{eqnarray}
 
 \item \underline{ $\epsilon_2$} : As defined above, $\epsilon_2$ 
 is related to the variation of the KR field energy density and thus to the field equation of the Kalb-Ramond field. However, the KR field energy density 
 in $F(R)$ ($\rho_{KR}$) and in the corresponding scalar-tensor theory ($\tilde{\rho}_{KR}$) are connected by 
 $\rho_{KR} = e^{-2\sigma}\tilde{\rho}_{KR}$ (with $\sigma = \sqrt{\frac{2}{3}}\kappa\xi$). Differentiating both sides of this expression 
 (with respect to $\tau$), one gets
 
 \begin{eqnarray}
  \frac{d\rho_{KR}}{d\tau}&=&\frac{d}{dt}\big[e^{-2\sigma}\tilde{\rho}_{KR}\big] \frac{dt}{d\tau}\nonumber\\
  &=&\rho_{KR} e^{-\sigma/2}\bigg[\frac{1}{\tilde{\rho}_{KR}}\frac{d\tilde{\rho}_{KR}}{dt} - 2\frac{d\sigma}{dt}\bigg]\ ,
  \label{time variation 1}
 \end{eqnarray}
 where we have used the relation among $\tau$ and $t$ as shown in eqn.~(\ref{cosmic time1 F(R)}). Recall that the evolution of $\tilde{\rho}_{KR}$ 
 (see eqn.~(\ref{evolution of KR energy density})) is given by,
 \begin{eqnarray}
  \frac{1}{\tilde{\rho}_{KR}}\frac{d\tilde{\rho}_{KR}}{dt} + 6H = 0\ .
  \label{time variation 2}
 \end{eqnarray}
 With the help of the expression (\ref{time variation 1}), the above equation can be written in terms of $\rho_{KR}$ as follows,
 \begin{eqnarray}
  \frac{\rho_{KR}'}{2\rho_{KR}} + e^{-\sigma/2} \dot{\sigma} + 3e^{-\sigma/2}H = 0\ ,
  \label{time variation 3}
 \end{eqnarray}
where prime and dot represent the derivative with respect to $\tau$ and $t$ respectively. Eqn.~(\ref{time variation 3}) along with the expression 
of $H_F$ (see eqn.(\ref{hubble F(R)})) lead to the final form of $\epsilon_2$ as follows
\begin{eqnarray}
 \epsilon_2&=&\frac{\rho_{KR}'}{2\rho_{KR}H_F}\nonumber\\
 &=&-3 + \frac{\dot{\sigma}}{2H_F}e^{-\sigma/2}\ .
 \label{second slow roll parameter}
\end{eqnarray}
\item \underline{$\epsilon_3$}: Using $F(R) = R + \frac{R^2}{m^2}$ (as we consider in the present context), $\epsilon_3$ can be 
simplified as,
\begin{eqnarray}
 \epsilon_3&=&\frac{1}{2F'(R)H_F}\frac{dF'(R)}{d\tau}\nonumber\\
 &=&\frac{1}{2RH_F}\frac{dR}{d\tau}\ ,
 \label{third 1}
\end{eqnarray}
where we consider $1 + \frac{2R}{m^2} \simeq \frac{2R}{m^2}$ near the beginning of inflation (as $n_s$ and $r$ are calculated at the onset of inflation). 
Furthermore, for a flat FRW metric, the Ricci scalar takes the form $R\simeq 12H_F^2$, and we get the final form of $\epsilon_3$ as,

\begin{eqnarray}
 \epsilon_3&=&\frac{1}{H_F^2}\frac{dH_F}{d\tau} = -\epsilon_F\nonumber\\
 &=&-\bigg[\frac{\frac{3\kappa^2h_0}{2a^6} + \frac{m^2}{12}e^{\sigma}(1 - e^{\sigma})}
  {\frac{m^2}{8}(1 - e^{\sigma})^2 + \frac{\kappa^2h_0}{2a^6} + \frac{m^2}{6}e^{\sigma}(1 - e^{\sigma})}\bigg]\ .
 \label{third slow roll paramater}
\end{eqnarray}

\item \underline{$\epsilon_4$}: As mentioned above, $E$ is defined as $E = \frac{\varTheta F'(R)H_F^2}{\rho_{KR}}$. Differentiating this expression (with respect to $\tau$), one gets,
\begin{eqnarray}
 \frac{E'}{EH_F} = \frac{\varTheta'}{\varTheta H_F} + \frac{1}{F'(R)H_F}\frac{dF'(R)}{d\tau} + 2\frac{H_F'}{H_F^2} - \frac{\rho_{KR}'}{\rho_{KR}H_F}\ .
 \label{fourth 1}
\end{eqnarray}
The above expression can be further simplified with the help of eqns.~(\ref{second slow roll parameter}) and (\ref{third slow roll paramater}),
\begin{eqnarray}
 \frac{E'}{EH_F} = \frac{\varTheta'}{\varTheta H_F} - 4\epsilon_F + 6 - \frac{\dot{\sigma}}{H_F}e^{-\sigma/2}\ .
 \label{fourth 2}
\end{eqnarray}
At this stage, we should obtain $\varTheta$ in order to get the final expression for $\epsilon_4$ as well as for $n_s$. 
By its definition, $\varTheta$ is given by,
\begin{eqnarray}
 \varTheta&=&\frac{\rho_{KR}}{H_F^2} + \frac{3}{2\kappa^2F'(R)H_F^2}\bigg(\frac{dF'(R)}{d\tau}\bigg)^2\nonumber\\
 &=&\frac{\rho_{KR}}{H_F^2} + 6\epsilon_F^2\big(\frac{F'(R)}{\kappa^2}\big)\ .
 \label{fourth 3}
\end{eqnarray}
In the second line of the above expression, we have used eqn.~(\ref{third slow roll paramater}). Differentiating both sides of eqn.~(\ref{fourth 3}), 
the following expression yields,
\begin{eqnarray}
 \frac{\varTheta'}{\varTheta H_F} = -2\epsilon_F + 2\frac{\epsilon_F'}{\epsilon_F H_F} + \frac{\kappa^2\rho_{KR}}{6F'(R)\epsilon_F^2 H_F^3} 
 \bigg[-6H_F + e^{-\sigma/2}\dot{\sigma} - \frac{H_F'}{H_F}\bigg]\ ,
 \label{fourth 4}
\end{eqnarray}
where we have used eqns.~(\ref{second slow roll parameter}) and (\ref{third slow roll paramater}). However, the above expression together with 
eqn.~(\ref{fourth 2}) lead to the final form of $\epsilon_4$ as
\begin{eqnarray}
 \epsilon_4&=&-3\epsilon_F + \frac{\epsilon_F'}{\epsilon_F H_F} + 3 - \frac{\dot{\sigma}}{2H_F}e^{-\sigma/2}\nonumber\\
 &+&\frac{\kappa^2\rho_{KR}}{6F'(R)\epsilon_F^2 H_F^3} \bigg[-6H_F + e^{-\sigma/2}\dot{\sigma} - \frac{H_F'}{H_F}\bigg]\ .
 \label{fourth slow roll parameter}
\end{eqnarray}
\end{itemize}

Hence, we can now calculate the spectral index. Introducing the above expressions of $\epsilon_{i}$ ($i = 1,2,3,4$ see 
eqns.(\ref{first slow roll parameter}), (\ref{second slow roll parameter}), (\ref{third slow roll paramater}) and (\ref{fourth slow roll parameter})) 
into eqn.(\ref{spectral index1}), we finally obtain the following form of $n_s$ as,
\begin{eqnarray}
 n_s = 1 - 2\frac{\epsilon_F'}{\epsilon_F H_F} 
 + \frac{\kappa^2\rho_{KR}}{6F'(R)\epsilon_F^2 H_F^3} \bigg[-6H_F + e^{-\sigma/2}\dot{\sigma} - \frac{H_F'}{H_F}\bigg]\ .
 \label{spectral index intermediate}
\end{eqnarray}

Note that in absence of the Kalb-Ramond field (i.e $\rho_{KR}=0$), $n_s$ turns out 
$n_s = 1 - 2\frac{\epsilon_F'}{\epsilon_F H_F}$, in agreement with the expression of spectral index in a pure $F(R)$ gravity model \cite{1}. 
However,l due to the presence of the KR field, $n_s$ is modified by the terms proportional to $\rho_{KR}$. Taking these modifications into account, 
the final form of $n_s$ is given by,

\begin{eqnarray}
 n_s = 1&-&2\bigg[\frac{-\frac{\kappa^2h_0}{2m^2} + \frac{1}{9}\bigg(\frac{e^{2\sigma_0}(1-e^{\sigma_0})(1-2e^{\sigma_0})}
 {(1-e^{\sigma_0})^2 + \frac{4\kappa^2h_0}{m^2}}\bigg)}{\frac{3\kappa^2h_0}{2m^2} + \frac{1}{12}e^{\sigma_0}(1-e^{\sigma_0})}\bigg] 
 + 2\bigg[\frac{-\frac{3\kappa^2h_0}{m^2} - \frac{1}{9}\bigg(\frac{e^{2\sigma_0}(1-e^{2\sigma_0})}
 {(1-e^{\sigma_0})^2 + \frac{4\kappa^2h_0}{m^2}}\bigg)}{\frac{\kappa^2h_0}{2m^2} + \frac{1}{8}(1-e^{\sigma_0})^2 
 + \frac{1}{6}e^{\sigma_0}(1-e^{\sigma_0})}\bigg]\nonumber\\ 
 &+&\frac{9}{16}\frac{\kappa^2h_0/m^2}
 {\bigg[\frac{1}{8}(1 - e^{\sigma_0})^2 + \frac{\kappa^2h_0}{2m^2} + \frac{1}{6}e^{\sigma_0}(1 - e^{\sigma_0})\bigg]
 \bigg[\frac{3\kappa^2h_0}{2m^2} + \frac{1}{12}e^{\sigma_0}(1 - e^{\sigma_0})\bigg]}\nonumber\\
 &-&\frac{3}{8}\frac{\kappa^2h_0/m^2}{\bigg[\frac{3\kappa^2h_0}{2m^2} + \frac{1}{12}e^{\sigma_0}(1 - e^{\sigma_0})\bigg]^2}\ .
 \label{spectral index2}
\end{eqnarray}

Let us now calculate the tensor to scalar ratio ($r$) which is defined as $r = 8\kappa^2 \frac{\varTheta}{F'(R)}$. By using 
the explicit expression of $\varTheta$, one gets
\begin{eqnarray}
 r = 8\kappa^2 \frac{\rho_{KR}}{H_F^2 F'(R)} + 48 \bigg(\frac{1}{2H_F F'(R)}\frac{dF'(R)}{d\tau}\bigg)^2\bigg|_{\tau_0}\ .
 \label{fifth 1}
\end{eqnarray}
We need the terms of the right hand side of eqn.~(\ref{fifth 1}) which can be obtained as follows:
\begin{itemize}
 \item \underline{First term in the r.h.s of eqn.~(\ref{fifth 1})}:\\
  With $F'(R) = 1 + 2R/m^2 \simeq 24H_F^2/m^2$ along with the expression $H_F = e^{-\sigma/2}\big[H + \frac{\dot{\sigma}}{2}$ 
 (see eqn.(\ref{hubble F(R)})), one gets,
 \begin{eqnarray}
  8\kappa^2 \frac{\rho_{KR}}{H_F^2 F'(R)}\bigg|_{\tau_0} = \frac{\kappa^2h_0}{\frac{3}{m^2}\bigg[H^2 + H\dot{\sigma}\bigg]^2}\ .
  \label{fifth 2}
 \end{eqnarray}
 
 By using the slow roll field equations, the above expression can be further simplified leading to the following form:
 \begin{eqnarray}
  8\kappa^2 \frac{\rho_{KR}}{H_F^2 F'(R)}\bigg|_{\tau_0} = \frac{3\kappa^2h_0/m^2}
  {\bigg[\frac{1}{8}(1 - e^{\sigma_0})^2 + \frac{\kappa^2h_0}{2m^2} + \frac{1}{6}e^{\sigma_0}(1 - e^{\sigma_0})\bigg]^2}\ .
  \label{fifth 3}
 \end{eqnarray}
 
 \item \underline{Second term in the r.h.s of eqn.(\ref{fifth 1})}:\\
  As obtained in eqn.(\ref{third slow roll paramater}), the second term of R.H.S of eqn.(\ref{fifth 1}) is equal to $48\epsilon_F^2$.
\end{itemize}
Hence, we obtain the tensor to scalar ratio:
\begin{eqnarray}
 r = \frac{3\kappa^2h_0/m^2}
  {\bigg[\frac{1}{8}(1 - e^{\sigma_0})^2 + \frac{\kappa^2h_0}{2m^2} + \frac{1}{6}e^{\sigma_0}(1 - e^{\sigma_0})\bigg]^2} + 48\epsilon_F^2\ .
  \label{ratio intermediate}
\end{eqnarray}

\begin{figure}[!h]
\begin{center}
 \centering
 \includegraphics[width=4.0in,height=3.8in]{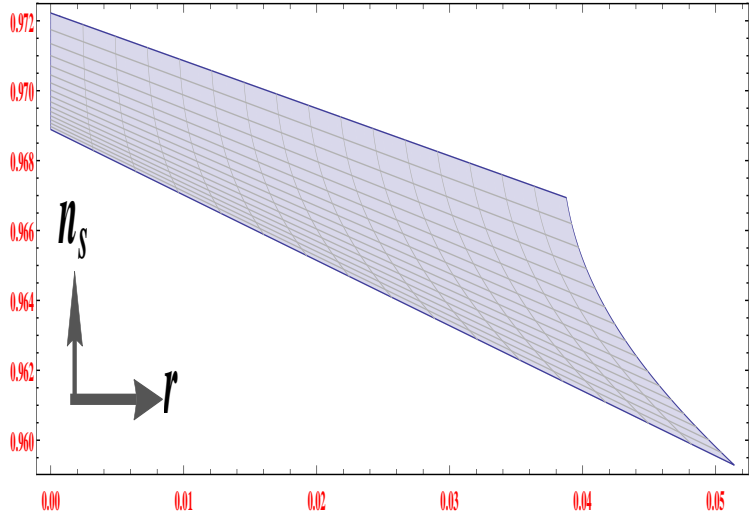}
 \caption{$n_s$ vs $r$ for $10 \leq |\sigma_0| \leq 14$ and $0.003 \leq \frac{\kappa^2h_0}{m^2} \leq 0.004$.}
 \label{plot_observable_4d}
\end{center}
\end{figure}

Note that from eqn.~(\ref{ratio intermediate}), whether $\rho_{KR} = 0$ (or equivalently $h_0 = 0$ i.e without the KR field), $r$ goes to $48\epsilon_F^2$ - the 
expression for tensor to scalar ratio in a pure $F(R)$ gravity model \cite{1}. However taking the effect of Kalb-Ramond field into account and substituting 
the expression of $\epsilon_F$ (obtained in eqn.~(\ref{slow roll parameter F(R)2})) in eqn.(\ref{ratio intermediate}), we get the final form 
of $r$ as follows:
\begin{eqnarray}
  r = \frac{3\frac{\kappa^2h_0}{2m^2} + 48\bigg[\frac{3\kappa^2h_0}{2m^2} + \frac{1}{12}e^{\sigma_0}(1 - e^{\sigma_0})\bigg]^2}
  {\bigg[\frac{1}{8}(1 - e^{\sigma_0})^2 + \frac{\kappa^2h_0}{2m^2} + \frac{1}{6}e^{\sigma}(1 - e^{\sigma})\bigg]^2}\ .
  \label{ratio2}
 \end{eqnarray}
 Thereby the final expressions of $n_s$ and $r$ are shown in eqns.~(\ref{spectral index2}) and (\ref{ratio2}) respectively, from which, it is evident 
 that both quantities depend on the parameters $\frac{\kappa^2h_0}{m^2}$ and $\sigma_0$. 
Eqns.~(\ref{spectral index2}) and (\ref{ratio2}) lead to the parametric 
 plot for $n_s$ vs. $r$ (with respect to the parameters $\frac{\kappa^2h_0}{m^2}$ and $\sigma_0$ ), as shown in  
 figures \ref{plot_observable_4d}.
 
 However,  observations based on Planck $2018$ impose a constraint on $n_s$ and $r$ as 
 $n_s = 0.9650 \pm 0.0066$ and $r < 0.07$ (combining with BICEP2/Keck - Array) respectively. 
 Therefore, figure \ref{plot_observable_4d} clearly indicates that for $|\sigma_0| > 10$ and 
 $0.003 < \frac{\kappa^2h_0}{m^2} < 0.004$, the theoretical values of $n_s$, $r$ (in the present context) match with the observational constraints. 
In addition, by the estimated values of $\frac{\kappa^2h_0}{m^2}$ and $\sigma_0$ ($\simeq -10$), the duration of inflation 
($\tau_f-\tau_0$, see eqn.(\ref{cosmic time2 F(R)})) becomes 
 $10^{-12}$(Gev)$^{-1}$ if the mass parameter ($m$) is separately taken as $10^{-5}$ (in Planckian units). We also obtain the number of 
e-foldings, defined by $N = \int_{0}^{\vartriangle \tau}H_F d\tau$ ($\vartriangle \tau = \tau_f-\tau_0$, duration of inflation), 
numerically, leadint to $N \simeq 56$ (with $\sigma_0 = -10$). These results are summarized in Table \ref{Table-1}.

\begin{figure}[!h]
\begin{center}
 \centering
 \includegraphics[width=3.5in,height=2.0in]{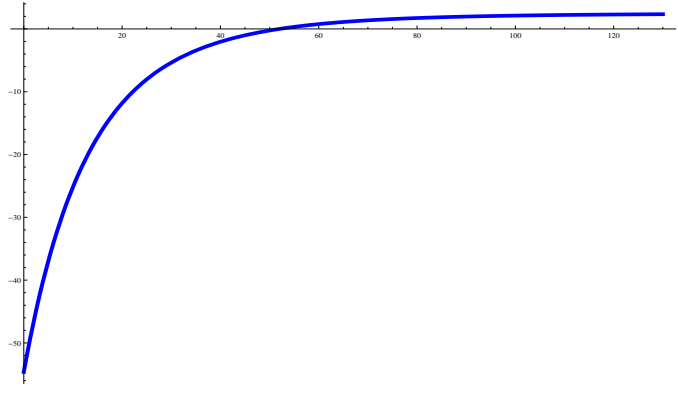}
 \caption{$q(\tau)$ vs $\tilde{\tau}$ for $\frac{\kappa^2h_0}{m^2} = 0.0035$, $\sigma_0 = -10$ and $m = 10^{-5}$ (in Planckian unit).}
 \label{plot_deceleration_parameter}
\end{center}
\end{figure}

\begin{table}[!h]
 \centering
  \begin{tabular}{|c| c|}
   \hline \hline
   Parameters & Estimated values\\
   \hline
   $n_s$ & $\simeq 0.9630$\\ 
   $r$ & $\simeq 0.03$\\
   $\tau_f-\tau_0$ & $10^{-12}$(GeV)$^{-1}$\\
   $N$ & 56\\
   \hline
  \end{tabular}%
  \caption{Estimated values of various quantities for $\frac{\kappa^2h_0}{m^2} = 0.0035$, $\sigma_0 = -10$ and $m = 10^{-5}$ (in Planckian unit)}
  \label{Table-1}
 \end{table}
 Table \ref{Table-1} clearly indicates that the present model may well explain the inflationary scenario of the universe 
 in terms of the observable quantities $n_s$ and $r$ as based on the results of Planck 2018.\\
 Using the solutions of $s(\tau)$ (see eqn. (\ref{scale factor2 F(R)}) along with the estimated values 
 of the parameters ($\frac{\kappa^2h_0}{m^2}$, $\sigma_0$, $m$), we depict the deceleration 
 parameter $q=-\frac{1}{s}\frac{d^2s}{d\tau^2}$ versus a dimensionless time variable $\tilde{\tau} = \frac{\tau}{\tau_f}N$ in Fig.~\ref{plot_deceleration_parameter}.\\

 Fig.~\ref{plot_deceleration_parameter} shows that the early universe starts from an accelerating stage with a graceful 
 exit at a finite time. However, from table [\ref{Table-1}], the maximum value for the parameter $\frac{\kappa^2h_0}{m^2}$ is given by 
 $\frac{\kappa^2h_0}{m^2} \simeq 0.004$, in order to match the present model with the observations of Planck 2018. 
 Taking $m=10^{-5}$ (in Planckian unit), we obtain $h_0^{max} \sim 10^{63}$ (GeV)$^{4}$. Recall that the term $h_0e^{[-2\sigma_0]}$ (see eqn.~(\ref{redefine})) 
 denotes the energy density for the KR field ($\rho_{KR}$) during the early universe in the $F(R)$ model. Therefore, the present model along with the constraints of Planck 2018 gives 
 an upper bound on the KR field energy density during the early universe as $\big(\rho_{KR}\big)^{max} \sim 10^{70}$ (GeV)$^{4}$ (with $\sigma_0 = -10$). In such situation, it is important 
 to examine whether the energy density of KR field (starting with $\sim 10^{70}$ (GeV)$^{4}$ from the early universe) 
 get suppressed and leads to a negligible footprint during our present universe. This matter is discussed in the next section.\\ 
 However, let us discuss briefly the cosmological evolution and the corresponding observable parameters 
 for the cases: (1) quadratic curvature gravity in absence of the Kalb-Ramond field (i.e for the pure $F(R)=R+R^2/m^2$ model),  
 (2) in the absence of higher order terms in the gravitational action, i.e for Einstein gravity with a KR field, and (3) when considering cubic curvature gravity with a KR field.
 \begin{enumerate}
  \item \underline{Quadratic curvature gravity in absence of the KR field}. In this case, the action of the model becomes,
  \begin{eqnarray}
   S = \int d^4x \sqrt{-g}\frac{1}{2\kappa^2}\bigg[R + \frac{R^2}{m^2}\bigg]\ .
 \label{new1}
  \end{eqnarray}
  
  Recall that $\frac{1}{2\kappa^2}=M_{p}^2$ (where $M_{p}$ is the four dimensional Planck mass). For this action, the solution of the FRW scale factor 
  can be obtained by fixing $h_0=0$ in the expression obtained in eqn.~(\ref{scale factor2 F(R)}), yielding
  \begin{eqnarray}
 s(\tau(t))&=&D\bigg(1 - \frac{2m(t-t_0)}{3\sqrt{6}C}\bigg)^{3/4}\nonumber\\ 
 &\exp&{\bigg[\frac{m(t-t_0)}{2\sqrt{6}} + \ln{\bigg(\frac{9}{-\sqrt{6}m(t-t_0) + 9C}\bigg)}\bigg]}\ ,
 \label{new2}
  \end{eqnarray}
  where $C = e^{-\sigma_0}$, $\tau$ is the cosmic time related to $t$ by eqn.~(\ref{cosmic time2 F(R)}) with $h_0=0$. 
Eqn.(\ref{new2}) leads to the acceleration of the early universe as,
\begin{eqnarray}
 \frac{1}{s}\frac{d^2s}{d\tau^2}\bigg|_{\tau\rightarrow \tau_0} = C \big(\frac{m}{2\sqrt{6}}\big)^2 (1-e^{-|\sigma_0|})^2\ ,
 \label{new3}
\end{eqnarray}
which clearly indicates that the early universe undergoes through an inflationary stage, the well known Starobinsky inflation. 
Correspondingly, the observational parameters as the spectral index and the tensor to scalar ratio depend only on the parameter $\sigma_0$ in absence of KR field. 
By introducing $h_0 = 0$ into the eqns.~(\ref{spectral index2}) and (\ref{ratio2}), 
one obtains the variation of $n_s$ and $r$ in terms of the parameter $\sigma_0$, as illustrated in figure \ref{plot_starobinsky}, which clearly shows that in absence of the 
Kalb-Ramond field, the spectral index and tensor to scalar ratio 
lie within the observational constraints for the interval $-5.0 \lesssim \sigma_0 \lesssim -4.5$. However, as mentioned above, even in the presence of the KR field, $n_s$ 
and $r$ also remain within the constraints but with a bound given by 
$\big(\rho_{KR}\big)^{max} \sim 10^{70}$ (GeV)$^{4}$ (with $\sigma_0 = -10$). For clearness, below we illustrate the comparison with/without the antisymmetric KR field in 
Table \ref{Table-comparison}.

 \begin{figure}[!h]
\begin{center}
 \centering
 \includegraphics[width=2.0in,height=2.0in]{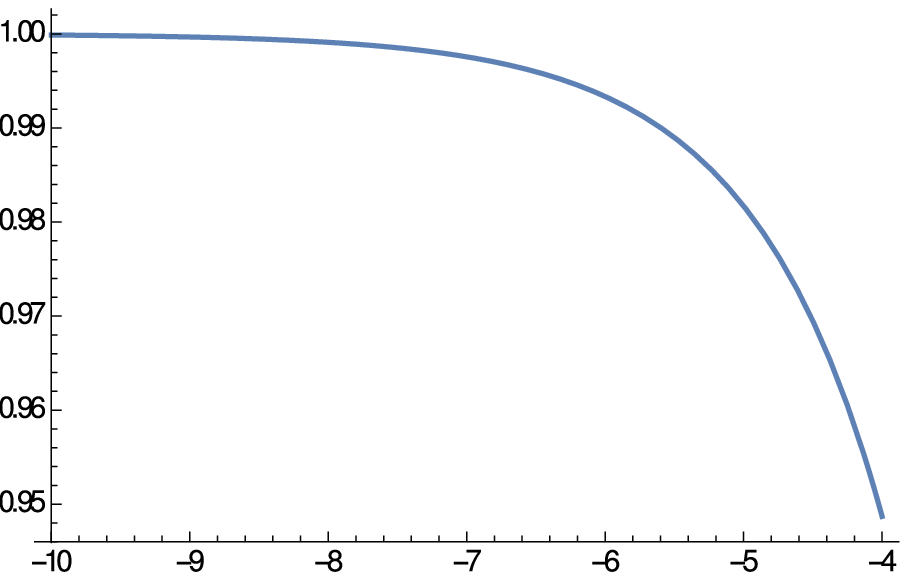}
 \includegraphics[width=2.0in,height=2.0in]{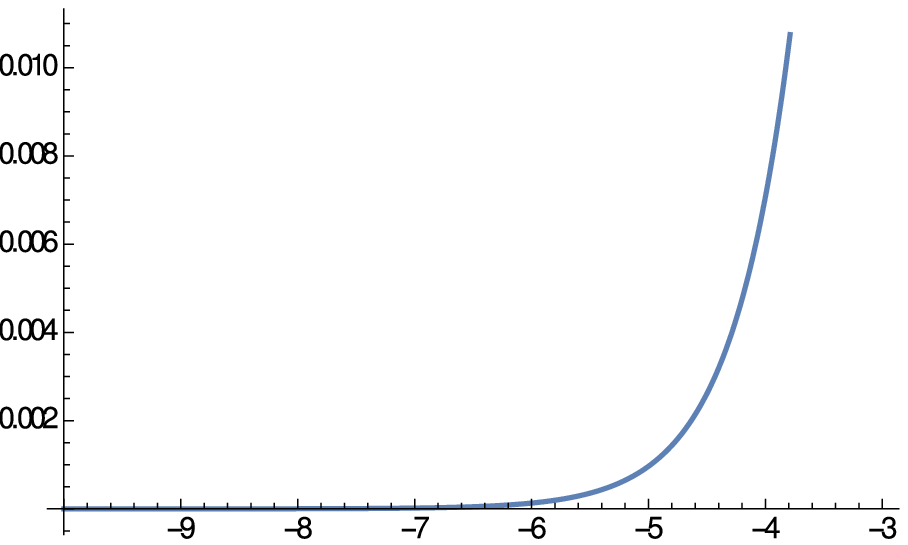}
 \caption{$n_s$ vs $\sigma_0$ (left panel), $r$ vs $\sigma_0$ (right panel).}
 \label{plot_starobinsky}
\end{center}
\end{figure}

\begin{table}[!h]
 \centering
  \begin{tabular}{|c| c| c|}
   \hline \hline
   Parameters & $\frac{\kappa^2h_0}{m^2}=0.0035$, $\sigma_0=-10$ & $h_0=0$, $\sigma_0=-4.5$\\
   \hline
   $n_s$ & $\simeq 0.9630$ & $\simeq 0.9660$\\ 
   \hline
   $r$ & $\simeq 0.03$ & $\simeq 0.003$\\
   \hline
  \end{tabular}%
  \caption{Comparison of $n_s$ and $r$ with/without the Kalb-Ramond field}
  \label{Table-comparison}
 \end{table}
 
 \item \underline{In absence of higher order curvature gravity}:
 Without higher order curvature degrees of freedom, the action takes the following form,
 \begin{eqnarray}
  S&=&\int d^4x \sqrt{-g}\bigg[\frac{R}{2\kappa^2} - \frac{1}{12}H_{\mu\nu\rho}H^{\mu\nu\rho}\bigg]\ .
 \label{new4}
 \end{eqnarray}

 As mentioned above, for a flat FLRW metric, the KR three tensor has only one non-zero component, i.e $H_{123}$ symbolized by $h_4$. With 
 this non-zero component, the pressure and energy density of the KR field turn out to be same and equal to $\frac{1}{2}h_4h^4$. As a consequence, 
 the FLRW equation becomes  $\dot{H} = -3H^2$ which can be solved and the scale factor yields $a(t) = (t-t_0)^{1/3}$. The acceleration of the 
 scale factor turns out negative and inflation does not occur. This result is in agreement with \cite{1808.04315}, which states that 
 a minimal model with an antisymmetric tensor field (in the Einstein frame) is not consistent with inflation.\\
 However, authors from \cite{1808.04315} showed that a stable de-Sitter solution can be achieved in the context of antisymmetric tensor field 
 by introducing a non-minimal coupling between the Ricci scalar and the tensor field. On the other hand, in the present paper, 
 we argue that the minimal prescription (in the presence of an antisymmetric tensor field) can also give rise to an inflationary era, but in the regime 
 of higher order curvature gravity.
 
 \item \underline{Cubic curvature gravity with the presence of the KR field}:
 In this case, the action is given by,
 \begin{eqnarray}
  S&=&\int d^4x \sqrt{-g}\bigg[\frac{1}{2\kappa^2}\big[R + \beta R^3\big] - \frac{1}{12}H_{\mu\nu\rho}H^{\mu\nu\rho}\bigg]\ .
 \label{cubic1}
 \end{eqnarray}
 Here $\beta$ is a free parameter with mass dimension [-4]. It is well known that $F(R) = R+\beta R^3$ does not give a {\it good inflation} i.e. 
 the theoretical values of $n_s$ and $r$ do not support the observable constraints from $Planck$ 2018. However, in the presence of an antisymmetric Kalb-Ramond 
 field, model (\ref{cubic1}) is consistent with $Planck$ 2018 constraints (i.e 
 $n_s = 0.9650 \pm 0.0066$ and $r < 0.07$, combining with BICEP2/Keck - Array). Here we present the plot for simultaneous compatibility of $n_s$, $r$ in Fig.~\ref{plot_cubic}:
  \begin{figure}[!h]
\begin{center}
 \centering
 \includegraphics[width=2.8in,height=3.0in]{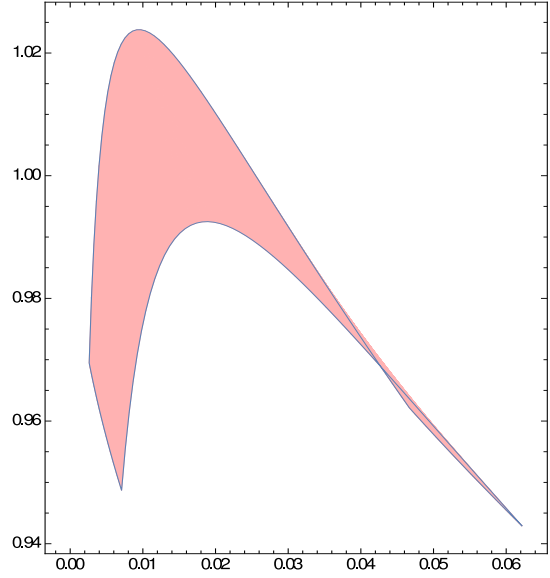}
 \caption{$n_s$ vs $r$ for $-5 \lesssim \sigma_0 \lesssim -4$ and $0.03 \lesssim \kappa^2h_0\sqrt{\beta} \lesssim 0.3$.}
 \label{plot_cubic}
\end{center}
\end{figure}

In Fig.~\ref{plot_cubic}, $e^{-\sigma_0} = 1+3\beta R_0^2$ (with $R_0$ be the spacetime curvature at the time of horizon crossing). In addition, Fig.~\ref{plot_cubic} clearly reveals 
that the observable parameters $\{ n_s, r\}$ remain within the confident regions provided by $Planck$ 2018.
 \end{enumerate}

 \subsubsection{Suppression of the Kalb-Ramond field in $F(R)$ gravity}
 
\begin{figure}[!h]
\begin{center}
 \centering
 \includegraphics[width=3.5in,height=2.0in]{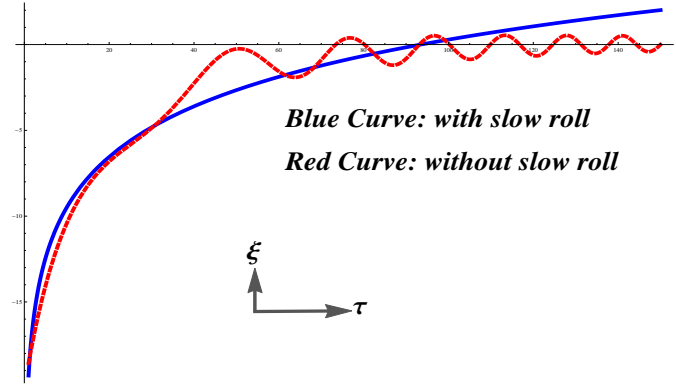}
 \caption{$\xi$ vs $\tilde{\tau}$ for $\frac{\kappa^2h_0}{m^2} = 0.0035$, $\sigma_0 = -10$ and $m = 10^{-5}$ (in Planckian unit).}
 \label{plot_field}
\end{center}
\end{figure}

\begin{figure}[!h]
\begin{center}
 \centering
 \includegraphics[width=3.5in,height=2.0in]{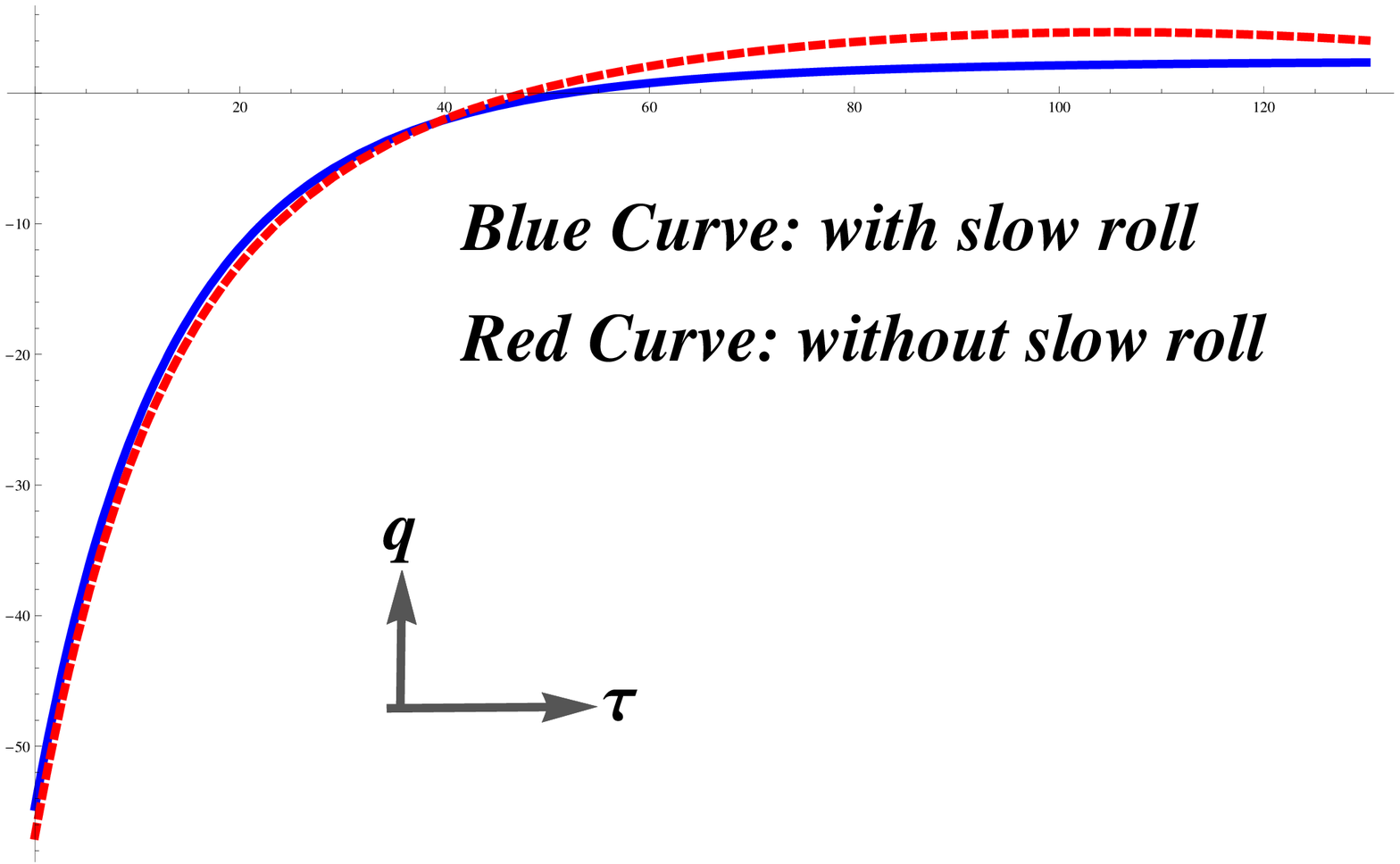}
 \caption{$q$ vs $\tilde{\tau}$ for $\frac{\kappa^2h_0}{m^2} = 0.0035$, $\sigma_0 = -10$ and $m = 10^{-5}$ (in Planckian unit).}
 \label{plot_deceleration_parameter1}
\end{center}
\end{figure}

\begin{figure}[!h]
\begin{center}
 \centering
 \includegraphics[width=3.5in,height=2.0in]{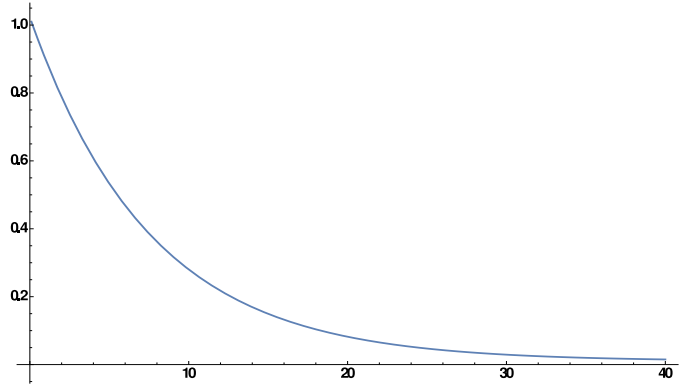}
 \caption{$\rho_{KR}$ vs $\tilde{\tau}$ for $\frac{\kappa^2h_0}{m^2} = 0.0035$, $\sigma_0 = -10$ and $m = 10^{-5}$ (in Planckian unit).}
 \label{plot_KR_field_suppression}
\end{center}
\end{figure}

 The energy density of the Kalb-Ramond field $\rho_{KR}$ in our $F(R)$ model in terms of the cosmic time $\tau$ is given by: 
  \begin{eqnarray}
  \rho_{KR}(\tau(t)) = e^{-2\sqrt{\frac{2}{3}}\kappa\xi(t)} \frac{h_0}{a^6}\ ,
  \label{energy density1}
 \end{eqnarray}
 where we have used the conformal transformation of the metric along with eqn.~(\ref{solution of KR energy density}). Therefore, 
 in order to address the effect of the KR field on our present universe, it is important to understand the late time evolution for $\xi(t)$ and $a(t)$. 
 As mentioned above, $\xi(t)$ goes to infinity at late times, starting from a negative value at the early universe. However, this 
 solution $\xi(t)$ is based on the slow-roll approximation which may not hold at late times. Then, let us relax the slow-roll 
 approximation, such that the field equations for $\xi(t)$ and $a(t)$ take the form:
  \begin{eqnarray}
 H^2 = \frac{\kappa^2}{3}\bigg[\frac{1}{2}\dot{\xi}^2 + \frac{m^2}{8\kappa^2}\big(1 - e^{\sqrt{\frac{2}{3}}\kappa\xi}\big)^2 
 + \frac{1}{2}\frac{h_0}{a^6}\bigg]\ ,
 \label{relax1}
\end{eqnarray}
and
\begin{eqnarray}
 \ddot{\xi} + 3H\dot{\xi} - \sqrt{\frac{2}{3}}\frac{m^2}{4\kappa}e^{\sqrt{\frac{2}{3}}\kappa\xi}\big(1 - e^{\sqrt{\frac{2}{3}}\kappa\xi}\big) = 0\ .
 \label{relax2}
 \end{eqnarray}
By solving these equations numerically, the evolution of the KR field and the deceleration parameter are depicted in Figs.~\ref{plot_field} and \ref{plot_deceleration_parameter1} respectively,  
where we have used the relation of $\tau(t)$ given in eqn.~(\ref{cosmic time2 F(R)}). Fig.~\ref{plot_field} shows the evolution of $\xi(\tau)$ for both cases, when assuming the slow-roll 
approximation and when no approximation is assumed.  As shown, the evolution for $\xi(\tau)$ is very similar in both cases. After inflation, the acceleration term for $\xi(\tau)$ starts 
to contribute and as a result both solutions (with and without slow roll conditions) differ from each other. 
 Similar conclusion holds for the deceleration parameter. Moreover in the slow roll
approximation, $\xi(\tau)$ does not tend to a finite value asymptotically, 
but goes to infinity at late times, while in absence of the slow roll approximation, $\xi(\tau)$ 
moves towards $<\xi> = 0$ asymptotically, showing an oscillatory behaviour at late times.\\
By using these numerical solutions for $\xi(\tau)$ and  $a(\tau)$, the KR field energy density (\ref{energy density1}) is obtained in our $F(R)$ model, $\rho_{KR}$, as shown in 
Fig.~\ref{plot_KR_field_suppression}, where the energy density of the KR field gradually decreases with the cosmic time ($\tau$) and the decaying 
time scale $\tilde{\tau}=40$ is smaller than the exit time from inflation ($\tilde{\tau}=56$). This may well explain why the present 
universe does not show any footprint of the antisymmetric Kalb-Ramond field.\\

However, besides the four dimensional context, higher dimensions spacetimes may provide a natural solution to the hierarchy problem i.e., apparent mismatch between the fundamental 
scale and the electroweak symmetry breaking scale \cite{arkani,horava,RS}. In such higher dimensional models, unlike electromagnetic or 
other matter fields, Kalb-Ramond field do propagate through extra dimensions and thus have Kaluza-Klein modes. Further attempts to unify gravity and electromagnetism require the 
inclusion of Kalb-Ramond field in higher-dimensional theories \cite{kubyshin,german}, such that the KR field may become important in the context of extra dimensional models. 
In the following sections, we discuss such higher dimensional spacetimes.
 
\section{Kalb-Ramond field in five dimensions in $F(R)$ gravity}
\label{Model-II}

Let us now investigate the cosmological evolution for the Kalb-Ramond field when higher dimensional spacetimes are considered. In particular, here we consider the well 
known Randall-Sundrum (RS) braneworld model with the presence of a Kalb-Ramond field in the bulk. RS model consists of one extra 
spatial dimension. The bulk spacetime is AdS in nature and $S^1/Z_2$ orbifolded along the extra dimension where the 
orbifold fixed points are identified with two 3-branes. 
If $\varphi$ is taken to be the extra 
dimensional angular coordinate, then the branes are 
located at $\varphi = 0$ (hidden brane) and at $\varphi = \pi$ (visible brane) respectively while the latter is identified with our visible universe. 
However, in such a braneworld scenario, the stabilization of interbrane separation (also known as modulus or radion) is an important issue to address 
and for this purpose one needs an extra stabilizing agent which is able to generate a stable radion potential. Here, in the present context, we consider 
quadratic curvature term in the five dimensional action together with the Einstein-Hilbert term as the stabilizing agent. Moreover, it is well known that higher order curvature 
terms become relevant in the limit of large curvature. Thus the RS bulk geometry, 
where the curvature is of the order of the Planck scale,  such that higher order curvature terms have to be included in the action. Hence, 
the action of the model is given by,
\begin{eqnarray}
 S&=&\int d^4xd\varphi \sqrt{-G} \bigg[\frac{(R + \alpha R^2)}{2\kappa^2} - \Lambda + V_h\delta(\varphi) + V_v\delta(\varphi-\pi)-\frac{1}{12}H_{MNL}H^{MNL}\bigg]\ ,\nonumber\\
 &=&S_{g} + S_{KR}\ ,
 \label{5d actionII}
\end{eqnarray}
where $G$ is the determinant of the five dimensional metric $G_{MN}$ ($M$, $N$, whose indexes runs from 0 to 4 where 0 to 3 are reserved for brane coordinates), 
$\alpha$ is a constant parameter having mass dimension [-2] and $\frac{1}{2\kappa^2}=M^3$ (with $M$ be the 5 dimensional Planck mass), while $\Lambda$ ($<0$) symbolizes the bulk 
cosmological constant and $V_h$, $V_v$ are the brane tensions on hidden and visible brane respectively. Moreover $H_{MNL}=\partial_{[M}B_{NL]}$ denotes 
the field strength tensor for the KR field $B_{NL}$ propagating in the five dimensional spacetime. However, as being allowed to propagate 
in the extra dimension, the KR field can be decomposed into Kaluza-Klein (KK) modes which are obviously coupled with to the extra dimensional modulus field. 
The overlap of these KK wave functions with the visible brane actually determines the strength of the KR field in our visible universe. In such a situation, 
it is important to explore the effects of higher order curvature terms on the dynamics of the modulus field which in turn controls the evolution 
of the bulk Kalb-Ramond field. These issues are addressed here from the perspective of four dimensional effective theory. In the following two 
subsections, we determine the effective four dimensional action individually for 
$S_{g}=\int d^4xd\varphi \sqrt{-G} \bigg[\frac{(R + \alpha R^2)}{2\kappa^2} - \Lambda + V_h\delta(\varphi) + V_v\delta(\varphi-\pi)\bigg]$ 
and $S_{KR}=\int d^4xd\varphi \sqrt{-G} \bigg[-\frac{1}{12}H_{MNL}H^{MNL}\bigg]$ respectively.\\
 
\subsubsection{Four dimensional effective action for $S_{g}$}

\begin{figure}[!h]
\begin{center}
 \centering
 \includegraphics[width=3.2in,height=2.2in]{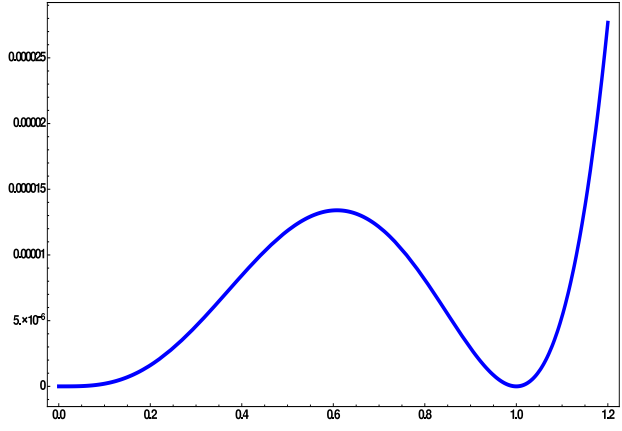}
 \caption{$U_{rad}(\Psi)$ vs $\Psi$ for $M = k = 1$ (in Planckian unit), $\omega=0.04$, $\kappa v_v = 10^{-7}$, $\frac{v_h}{v_v} = 1.2$ 
 and $\alpha = \frac{1}{M^2}$.}
 \label{plot_potential}
\end{center}
\end{figure}

In order to find the effective action of $S_g$, we need the solution for the five dimensional spacetime metric $G_{MN}$. For this purpose, 
first we determine the field(s) solutions in the corresponding scalar-tensor (ST) theory and then transform the solutions back to the Jordan frame by using the inverse 
conformal transformation. Following section \ref{sectionFR}, the conformal transformation of the spacetime metric can be expressed as 
$G_{MN}\rightarrow \tilde{G}_{MN} = e^{\frac{\kappa\Phi}{\sqrt{3}}}G_{MN}$, while the action $S_g$ leads to:
\begin{eqnarray}
 S_g[\Phi,\tilde{G}_{MN}]&=&\int d^4x d\varphi \sqrt{\tilde{G}} \bigg[\frac{\tilde{R}}{2\kappa^2} - \frac{1}{2}\tilde{G}^{MN}\partial_M\Phi 
 \partial_N\Phi - V(\Phi) - \Lambda\nonumber\\
 &-&e^{-\frac{5}{2\sqrt{3}}\kappa\Phi} V_h\delta(\varphi) - e^{-\frac{5}{2\sqrt{3}}\kappa\Phi} V_v\delta(\varphi-\pi)\bigg]\ ,
 \label{action1STII}
\end{eqnarray}
where $\Phi$ is the scalar field in ST theory and $V(\Phi)$ is its potential which takes the following form
\begin{eqnarray}
 V(\Phi)&=&\frac{1}{8\kappa^2\alpha} \exp{(-\frac{5}{2\sqrt{3}}\kappa\Phi)} 
 \bigg[\exp{(\frac{3}{2\sqrt{3}}\kappa\Phi)} - 1\bigg]^2\nonumber\\
 &-&\Lambda \bigg[\exp{(-\frac{5}{2\sqrt{3}}\kappa\Phi)} + 1\bigg]\ .
 \label{scalar_potentialII}
\end{eqnarray}
Moreover, the last two terms in eqn.~(\ref{action1STII}) are contributions from the brane tensions of hidden and visible branes. 
However, in order to check the stability of $V(\Phi)$, we take the single derivative with respect to $\Phi$
in both sides of eqn.~(\ref{scalar_potentialII}),
\begin{eqnarray}
 V'(\Phi) = \frac{1}{16\sqrt{3}\kappa\alpha} e^{-\frac{5}{2\sqrt{3}}\kappa\Phi} \bigg[e^{\frac{6}{2\sqrt{3}}\kappa\Phi} 
 + 4e^{\frac{3}{2\sqrt{3}}\kappa\Phi} - \bigg(5-40\kappa^2\alpha\Lambda\bigg)\bigg]\ ,
 \label{derivativeII}
\end{eqnarray}
which immediately leads to the fact that $V(\Phi)$ is stable only for $\alpha > 0$. Correspondingly the vacuum expectation value $<\Phi>$  and the 
squared mass ($m_{\Phi}^2$) of $\Phi$ are given by,
\begin{equation}
 \exp{\bigg(\frac{3}{2\sqrt{3}}\kappa<\Phi>\bigg)} = \bigg[\sqrt{9 - 40\kappa^2\alpha\Lambda} - 2\bigg]\ ,
 \label{vev_phiII}
\end{equation}
and
\begin{eqnarray}
 m_{\Phi}^2 = \frac{1}{8\alpha} \bigg[\sqrt{9 - 40\kappa^2\alpha\Lambda}\bigg] \bigg[\sqrt{9 - 40\kappa^2\alpha\Lambda} - 2\bigg]^{-\frac{2}{3}}\ .
 \label{mass_phiII}
\end{eqnarray}
 As we will see below the stability of the modulus field is also ensured by the condition $\alpha>0$ - same as for the stability 
of $V(\Phi)$. Thus, it can be argued that the stability of $V(\Phi)$ and of interbrane separation are intimately connected in the 
higher order curvature RS model. However, note that the minimum value of $V(\Phi)$ is non-zero and is given by,
\begin{eqnarray}
 V(<\Phi>)&=&\Lambda + [\sqrt{9 - 40\kappa^2\alpha\Lambda} - 2]^{-\frac{5}{3}}\nonumber\\
 &\bigg[&-\Lambda + (1/8\kappa^2\alpha)[\sqrt{9 - 40\kappa^2\alpha\Lambda}-3]^2\bigg]\ .
 \nonumber
\end{eqnarray}
This non-zero value of the potential works as a cosmological constant together with $\Lambda$ and thus the effective cosmological constant 
in ST theory is given by $\Lambda_{eff}=\Lambda+V(<\Phi>)$ (a simple algebra shows that $\Lambda_{eff}$ is negative). By considering a small fluctuation of the scalar field 
around its stable value as $\Phi = <\Phi> + \xi$, the action (\ref{action1STII}) can be written as follows,
\begin{eqnarray}
 S_g[\Phi,\tilde{G}_{MN}]&=&\int d^4x d\varphi \sqrt{\tilde{G}} \bigg[\frac{\tilde{R}}{2\kappa^2} - \frac{1}{2}\tilde{G}^{MN}\partial_M\xi 
 \partial_N\xi - (1/2)m_{\Phi}^2\xi^2 - \Lambda_{eff}\nonumber\\
 &-&e^{-\frac{5}{2\sqrt{3}}\kappa(<\Phi>+\xi)} V_h\delta(\varphi) - e^{-\frac{5}{2\sqrt{3}}\kappa(<\Phi>+\xi)} V_v\delta(\varphi-\pi)\bigg]\ ,
 \label{action2STII}
\end{eqnarray}
where we keep the terms up to quadratic order in $\xi$. As expected, the scalar-tensor action contains 
two independent fields: $\Phi$ and $\tilde{G}_{MN}$. Let us now find the corresponding solutions of the field equations. 
By assuming a negligible backreaction of the scalar field ($\Phi$) on the background spacetime, the metric $\tilde{G}_{MN}$ is given by the well known Randall-Sundrum solution as,
\begin{eqnarray}
 d\tilde{s}^2 = e^{- 2 kr_c|\varphi|} \eta_{\mu\nu} dx^{\mu} dx^{\nu} + r_c^2d\varphi^2 ,
 \label{grav.sol1.STII}
\end{eqnarray}
where $k = \sqrt{\frac{-\Lambda_{eff}}{24M^3}}$ and $r_c$ is the compactification radius 
of the extra dimension in ST theory. Moreover, the brane tensions are given by following expressions:
\begin{eqnarray}
 V_h = 24M^3k* \exp{\bigg[\frac{5}{2\sqrt{3}}\kappa(<\Phi>+v_h)\bigg]}\ ,\nonumber\\
 V_v = -24M^3k* \exp{\bigg[\frac{5}{2\sqrt{3}}\kappa(<\Phi>+v_v)\bigg]}\ .
 \nonumber
\end{eqnarray}
Here $v_h$ and $v_v$ are the boundary values of $\xi$ on the hidden and visible brane respectively. Together with the metric (\ref{grav.sol1.STII}), the scalar field $\Phi$ 
equation turns out to be,
\begin{eqnarray}
 -\frac{1}{r_c^2}\partial_\varphi[\exp{(-4kr_c|\varphi|)}\partial_\varphi\xi] + m_{\Phi}^2\exp{(-4kr_c|\varphi|)}\xi(\varphi) = 0\ ,
 \label{eom.scalar.fieldII}
\end{eqnarray}
where the scalar field $\xi$ is considered to be the function of an extra dimensional coordinate only. By taking non-zero values of $\xi$ on the branes, the above differential 
equation has the following solution,
\begin{eqnarray}
 \xi(\varphi) = e^{2kr_c|\varphi|} \big[Ae^{\nu kr_c|\varphi|} + Be^{-\nu kr_c|\varphi|}\big]\ ,
 \label{sol.scalar.fieldII}
\end{eqnarray}
with $\nu = \sqrt{4 + m_{\Phi}^2/k^2}$. Furthermore, $A$, $B$ are integration constants that can be obtained from the boundary conditions $\xi(0)=v_h$ and $\xi(\pi)=v_h$, as follows,
\begin{equation}
 A = v_v e^{-(2+\nu)kr_c\pi} - v_h e^{-2\nu kr_c\pi}\ ,
 \nonumber\\
\end{equation}
and
\begin{equation}
 B = v_h (1 + e^{-2\nu kr_c\pi}) - v_v e^{-(2+\nu)kr_c\pi}\ .
 \nonumber\\
\end{equation}
Thus, eqns.~(\ref{grav.sol1.STII}) and (\ref{sol.scalar.fieldII}) specify the field solutions in this spacetime. Recall that the original $F(R)$ model is represented by the action 
$S_{g}[G_{MN}]=\int d^4xd\varphi \sqrt{-G} \bigg[\frac{(R + \alpha R^2)}{2\kappa^2} - \Lambda + V_h\delta(\varphi) + V_v\delta(\varphi-\pi)\bigg]$. The solution of the 
spacetime metric ($G_{MN}$) in the original $F(R)$ model can be obtained from the solutions of the corresponding scalar-tensor 
theory with the help of the inverse conformal transformation. Thus, the line element turns out to be,
\begin{equation}
 ds^2 = e^{-\frac{\kappa}{\sqrt{3}}\Phi(\varphi)} \bigg[e^{- 2 kr_c|\varphi|} \eta_{\mu\nu} dx^{\mu} dx^{\nu} + r_c^2d\varphi^2\bigg]\ ,
 \label{grav.sol1.F(R)II}
\end{equation}
where $\Phi(\varphi) = <\Phi> + \xi(\varphi)$ and $\xi(\varphi)$ are obtained in eqn.~(\ref{sol.scalar.fieldII}). In order to introduce the radion field, $r_c$ is 
replaced by $T(x)$, known as radion (or modulus) field. For simplicity, here we consider that this new field depends only on the brane coordinates. Thus, the line element becomes,
\begin{eqnarray}
 ds^2 = e^{-\frac{\kappa}{\sqrt{3}}\Phi(x,\varphi)} \bigg[e^{- 2 kT(x)|\varphi|} 
 g_{\mu\nu}(x) dx^{\mu} dx^{\nu} + T(x)^2d\varphi^2\bigg]\ .
 \label{grav.sol2.F(R)}
\end{eqnarray}
Here $g_{\mu\nu}(x)$ is the induced on-brane metric and $\Phi(x,\varphi)$ can be obtained from (\ref{sol.scalar.fieldII}) by 
replacing $r_c$ by $T(x)$. Substituting the above solution of $G_{MN}$ into the action $S_{g}[G_{MN}]$ and integrating over the extra 
dimensional coordinate $\varphi$, the effective four dimensional on-brane action becomes:
\begin{eqnarray}
 A_{eff}^{1} = \int d^4x \sqrt{-g} \bigg[M_{(4)}^2R_{(4)} - \frac{1}{2}g^{\mu\nu}\partial_{\mu}\Psi\partial_{\nu}\Psi - U_{rad}(\Psi)\bigg]\ ,
 \label{effective action1II}
\end{eqnarray}
where $M_{(4)}^2= \frac{M^3}{k}\bigg[\sqrt{9 - 40\kappa^2\alpha\Lambda} - 2\bigg]^{1/2}$ is the 
four dimensional Planck scale, $R_{(4)}$ is the on-brane Ricci scalar formed by $g_{\mu\nu}(x)$. Moreover,  
$\Psi(x) = \sqrt{\frac{24M^3}{k}} \big[1 + \frac{20}{\sqrt{3}}\alpha k^2\kappa v_h\big] e^{-k\pi T(x)} = fe^{-k\pi T(x)}$ 
(with $f = \sqrt{\frac{24M^3}{k}} [1 + \frac{20}{\sqrt{3}}\alpha k^2\kappa v_h]$), is the 
canonical radion field and $U_{rad}(\Psi)$ is the radion potential with the following form \cite{tp1},
\begin{eqnarray}
 U_{rad}(\Psi) = \frac{20}{\sqrt{3}}\frac{\alpha k^5}{M^6} \Psi^4 
 \bigg[(v_h - \frac{\kappa v_h^2}{2\sqrt{3}} + \frac{\kappa v_hv_v}{2\sqrt{3}})(\Psi/f)^\omega - v_v\bigg]^2\ ,
 \label{potential.radion.F(R)II}
\end{eqnarray}
where the terms proportional to $\omega$ ($= \frac{m_{\Phi}^2}{k^2}<1$, which is also consistent with observational bounds) 
are neglected. Note that $U_{rad}(\Psi)$ goes to zero as the higher order curvature parameter $\alpha$ tends to zero. 
However, as $\alpha\rightarrow 0$, the action contains 
only the Einstein-Hilbert term which is not able to generate any potential for the modulus field, as shown in \cite{GW_radion}. 
Thereby, the potential term for the radion field is generated entirely 
due to the presence of the higher order curvature term ($\alpha R^2$) in the action. Hence, the sign of the higher 
curvature term comes through the radion potential in the four dimensional effective action. In this context, the stabilization of the interbrane separation is based on whether the 
radion potential is stable or not. For $\alpha>0$, the potential $U_{rad}$ has a minima and a maxima at
\begin{eqnarray}
 \Psi_{min}&=&<\Psi>=\bigg[\frac{v_v f^{\omega}}{\big(v_h - \frac{\kappa v_h^2}{2\sqrt{3}} + \frac{\kappa v_vv_h}{2\sqrt{3}}\big)}\bigg]^{1/\omega}\ ,
 \label{vev.radion.F(R)II}
\end{eqnarray}
and
\begin{eqnarray}
 \Psi_{max} = \bigg[\bigg(\frac{2}{2+\omega}\bigg)
 \bigg(\frac{v_vf^{\omega}}{v_h - \frac{\kappa v_h^2}{2\sqrt{3}} + \frac{\kappa v_vv_h}{2\sqrt{3}}}\bigg)\bigg]^{1/\omega}\ .
 \nonumber
\end{eqnarray}
The minima of $U_{rad}(\Psi)$ immediately leads to the stabilization of the interbrane separation,
\begin{eqnarray}
 k\pi <T(x)> = \frac{4k^2}{m_{\Phi}^2}[\ln{(\frac{v_h}{v_v})} - \frac{\kappa v_v}{2\sqrt{3}}(\frac{v_h}{v_v} - 1)]\ .
 \label{brane separationII}
\end{eqnarray}
The expression of $m_{\Phi}^2$ as obtained in eqn.~(\ref{mass_phiII}), clearly indicates that $<T(x)>$ is proportional to  
parameter $\alpha$. Thus, the model considered here would collapse as $\alpha$ tends to zero, as pointed out in the discussion above. 
Moreover, eqn.~(\ref{potential.radion.F(R)II}) imposes that 
$U_{rad}(\Psi)$ goes to zero at $\Psi=0$. In figure \ref{plot_potential},  the potential $U_{rad}(\Psi)$ is depicted.
 
\subsubsection{Effective action for $S_{KR}$}

Recall that the 5D KR field action is given by,
\begin{eqnarray}
 S_{H} = -\frac{1}{12}\int d^4xd\varphi \sqrt{-G}\bigg[H_{MNL}H^{MNL}\bigg]
 \label{KR action1}
\end{eqnarray}
where the KR field strength tensor $H_{MNL}$ is related to $B_{MN}$ (second rank antisymmetric tensor field) as 
$H_{MNL} = \partial_{[M}B_{NL]}$, with latin and greek indices running from $0$ to $4$ and $0$ to $3$ respectively. 
It is straightforward to see that the action $S[H]$ is invariant under the gauge transformation $B_{MN} \rightarrow B_{MN}+\partial_{[M}W_{N]}$, with $W_{N}$ as an 
arbitrary function of spacetime coordinates. This gauge invariance of the KR field allows us to set $B_{4\mu} = 0$. Then, by using the form of $G_{MN}$ and 
keeping $B_{4\mu}=0$, the 
above action turns out to be,
\begin{eqnarray}
 S_{H}&=&-\frac{1}{12} \int d^4xd\varphi \sqrt{-g}e^{2kT(x)\varphi}T(x) 
 \bigg[g^{\mu\alpha}g^{\nu\beta}g^{\lambda\gamma}H_{\mu\nu\lambda}H_{\alpha\beta\gamma}\nonumber\\
 &-&\frac{3}{T(x)^2}e^{-2kT(x)\varphi}g^{\mu\alpha}g^{\nu\beta}B_{\mu\nu}\partial_{\phi}^2B_{\alpha\beta}\bigg]\ .
 \label{KR action2}
\end{eqnarray}
The Kaluza-Klein decomposition for the KR field can be written as,
\begin{eqnarray}
 B_{\mu\nu}(x,\varphi) = \sum B_{\mu\nu}^{(n)}(x)\chi^{(n)}(x,\varphi)\ ,
 \label{KR decompose}
\end{eqnarray}
where $B_{\mu\nu}^{(n)}(x)$ and $\chi^{(n)}(x,\varphi)$ represent the $n$th mode of on-brane KR field and extra dimensional 
KR wave function respectively. The wave function $\chi^{(n)}$ 
is considered to be a function of the brane coordinates also (apart from the coordinate $\varphi$), as we are interested to investigate whether the dynamical evolution of the 
KR field leads to its invisibility in the present universe. \\

By substituting the decomposition in the 5-dimensional action $S_{H}$ and integrating over the extra dimension, the four dimensional effective action turns out to be:
\begin{eqnarray}
 A_{eff}^{(2)}&=&-\frac{1}{12} \int d^4x \sqrt{-g}
 \bigg[g^{\mu\alpha}g^{\nu\beta}g^{\lambda\gamma}H_{\mu\nu\lambda}^{(n)}H_{\alpha\beta\gamma}^{(n)}\nonumber\\
 &+&3m_n^2 g^{\mu\alpha}g^{\nu\beta}B_{\mu\nu}^{(n)}B_{\alpha\beta}^{(n)}\bigg]\ ,
 \label{effective action2II}
\end{eqnarray}
as far as $\chi^{(n)}(x,\phi)$ satisfies the following equation of motion, 
\begin{eqnarray}
 \frac{\partial\chi^{(n)}}{\partial t} \frac{\partial\chi^{(m)}}{\partial t} 
 - \frac{1}{T^2(t)}e^{-2kT(t)\varphi}\chi^{(n)}\frac{\partial^2\chi^{(m)}}{\partial\varphi^2} = m_n^2\chi^{(n)}\chi^{(m)}\ ,
 \label{wave function equation}
\end{eqnarray}
along with the normalization condition,
\begin{eqnarray}
 \int_0^{\pi} d\varphi e^{2kT(t)\varphi}\chi^{(n)}\chi^{(m)} = \frac{1}{T^2(t)}\delta_{mn}\ ,
 \label{wave function normalization}
\end{eqnarray}
where $m_n$ denotes the mass of nth KK mode. As we will see below, obtaining the coupling between the KR field and the Standard Model fields on the visible brane is important. 
Furthermore, eqn.(\ref{wave function equation}) clearly shows that the dynamical evolution of $\chi^{(n)}(x,\varphi)$ 
is coupled to the modulus (or radion) field $T(x)$. Eqns.~(\ref{effective action1II}) and (\ref{effective action2II}) immediately leads to the full form of four dimensional 
effective action as follows :
\begin{eqnarray}
 A_{eff}&=&A_{eff}^{(1)} + A_{eff}^{(2)}\nonumber\\
 &=&\int d^4x\sqrt{-g}\bigg[\frac{M^3}{k}R^{(4)} - \frac{1}{2}g^{\mu\nu}\partial_{\mu}\Psi\partial_{\nu}\Psi - U_{rad}(\Psi)\nonumber\\
 &-&\frac{1}{12}\bigg(g^{\mu\alpha}g^{\nu\beta}g^{\lambda\gamma}H_{\mu\nu\lambda}^{(n)}H_{\alpha\beta\gamma}^{(n)} 
 + 3m_n^2 B_{\mu\nu}^{(n)}B^{\mu\nu(n)}\bigg)\bigg]\ ,
 \label{full effective action_secondary II}
\end{eqnarray}
where $U_{rad}(\Psi)$ is explicitly shown in eqn.(\ref{potential.radion.F(R)II}). From now on, we deal with the zeroth Kaluza-Klein mode of Kalb-Ramond field for which $m_{n=0} = 0$. 
With this lowest KK mode, the four dimensional effective action turns out to be,
\begin{eqnarray}
 A_{eff}&=&A_{eff}^{(1)} + A_{eff}^{(2)}\nonumber\\
 &=&\int d^4x\sqrt{-g}\bigg[\frac{M^3}{k}R^{(4)} - \frac{1}{2}g^{\mu\nu}\partial_{\mu}\Psi\partial_{\nu}\Psi - U_{rad}(\Psi)\nonumber\\ 
 &-&\frac{1}{12}g^{\mu\alpha}g^{\nu\beta}g^{\lambda\gamma}H_{\mu\nu\lambda}^{(0)}H_{\alpha\beta\gamma}^{(0)}\bigg]\ ,
 \label{full effective action II}
\end{eqnarray}
Due to the presence of $U_{rad}(\Psi)$, the radion field acquires some certain dynamics which affect the dynamical evolution of the 
KR wave function $\chi^{(0)}(x,\varphi)$, as $\Psi$ and $\chi^{(0)}(x,\varphi)$ are coupled through the eqn.~(\ref{wave function equation})). 
In such a scenario, our motivation is to investigate whether the evolution of $\chi^{(0)}(x,\varphi)$ leads to a negligible footprint of the KR field in the present visible 
universe. However, it was shown earlier in \cite{sengupta4} that the effect of the KR field may be significant and can play an important role in the early era of the universe. 
Therefore, in order to address the dynamical suppression of the KR field, it is important to start from the very early universe where we will investigate whether the universe 
passes through an inflationary era. For these purposes, we try to solve the cosmological Friedmann equations obtained from $A_{eff}$ in the following sections. 

\subsection{Effective cosmological equations and solutions}

The on-brane metric ansatz that fits our purpose in the present context can be expressed as follows,
\begin{eqnarray}
 ds_{(4)}^2&=&g_{\mu\nu}(x)dx^{\mu}dx^{\nu}\nonumber\\
 &=&-dt^2 + b^2(t)\big[dx^2 + dy^2 + dz^2\big]\ ,
 \label{4d metricII}
\end{eqnarray}
where $b(t)$ is the scale factor of our universe. With this ansatz along with the expressions of the energy-momentum tensor for the
four dimensional KR field (as shown in previous section), we obtain the following Einstein's field equations for the action $A_{eff}$,
\begin{eqnarray}
 3H_b^2 = \frac{1}{2}\dot{\Psi}^2 + \frac{20}{\sqrt{3}}\frac{\alpha k^5}{M^6}v_v^2 \Psi^4 \bigg[F\Psi^\omega - 1\bigg]^2 + \frac{1}{2}h_4h^4\ ,
 \label{einstein equation1II}
\end{eqnarray}
\begin{eqnarray}
 2\dot{H}_b + 3H_b^2 + \frac{1}{2}\dot{\Psi}^2 - \frac{20}{\sqrt{3}}\frac{\alpha k^5}{M^6}v_v^2 \Psi^4 \bigg[F\Psi^\omega - 1\bigg]^2 
 + \frac{1}{2}h_4h^4 = 0
 \label{einstein equation2II}
\end{eqnarray}
where $H_{b}=\frac{\dot{b}}{b}$ is known as the on-brane Hubble parameter, 
$F = \frac{1}{v_vf^{\omega}}\big(v_h - \frac{\kappa v_h^2}{2\sqrt{3}} + \frac{\kappa v_vv_h}{2\sqrt{3}}\big)$ and 
$h_4 = H_{123}^{(0)}$ (as the other components 
of $H_{\mu\nu\lambda}^{(0)}$ vanishes as given by the off-diagonal Einstein's equations). Moreover, 
the field equations for $H_{\mu\nu\lambda}^{(0)}$ and for radion field ($\Psi$) are given by,
\begin{eqnarray}
 \nabla_{\mu}H^{\mu\nu\lambda(0)} = \frac{1}{\sqrt{-g}}\partial_{\mu}\bigg[\sqrt{-g}H^{\mu\nu\lambda(0)}\bigg] = 0\ ,
 \label{KR equationII}
\end{eqnarray}
and
\begin{eqnarray}
 \ddot{\Psi} + 3H\dot{\Psi} + \frac{80}{\sqrt{3}}\frac{\alpha k^5}{M^6}v_v^2 \Psi^3 \bigg[F\Psi^\omega - 1\bigg]^2 = 0\ .
 \label{radion equationII}
\end{eqnarray}
Following Appendix \ref{appendix-I}, eqn.~(\ref{KR equationII}) leads to a non-zero component of $H_{\mu\nu\lambda}^{(0)}$ i.e. $h_4$ 
depends on the cosmic time $t$, as also expected from the gravitational field equations. Differentiating both sides of (\ref{einstein equation1II}) with respect to $t$, the 
following expression is obtained,
\begin{eqnarray}
 6H_b\dot{H}_b = \dot{\Psi}\ddot{\Psi} + \frac{80}{\sqrt{3}}\frac{\alpha k^5}{M^6}v_v^2 \Psi^3 \bigg[F\Psi^\omega - 1\bigg]^2\dot{\Psi} + 
 \frac{1}{2}\frac{d}{dt}\big(h_4h^4\big)\ .
 \nonumber
\end{eqnarray}
Furthermore, eqns.~(\ref{einstein equation1II}) and (\ref{einstein equation2II}) immediately give $2\dot{H}_b = \dot{\Psi}^2 - \frac{1}{2}h_4h^4$. With this 
expression for $\dot{H}_b$ along with the above equation, we obtain the cosmic evolution for the energy density of the 
on-brane KR field ($\Omega_{KR} = \frac{1}{2}h_4h^4$) as $\frac{d}{dt}\Omega_{KR} = -6H_b \Omega_{KR}$. Solving this equation, we get
\begin{eqnarray}
 \Omega_{KR} = \frac{\Omega_{0}}{b^6}\ ,
 \label{energy densityII}
\end{eqnarray}
where $\Omega_0$ is an integration constant. Eqn.~(\ref{energy densityII}) clearly indicates that the on-brane KR field energy density 
is proportional to $1/b^6$ (same as previous model) and thus decreases with the expansion of the universe. Moreover, note 
that $\Omega_{KR}$ decreases more rapidly in comparison to normal matter ($\propto 1/b^3$) as well as radiation ($\propto 1/b^4$) energy density. 
This may well explain why the Kalb-Ramond field has negligible footprint in the present visible universe. However, at the same time, 
eqn.~(\ref{energy densityII}) also reveals that the KR field may has a significant contribution during the early universe (when $b(t)$ is small in comparison 
to the present one). On the other hand, recall that the bulk KR field has also an extra dimensional Kaluza-Klein (KK) wave 
function (besides the on-brane part) which determines the coupling among the on-brane KR field and other matter fields. Thereby, along with the on-brane 
part, the extra dimensional wave function $\chi^{(0)}(t,\varphi)$ also plays a crucial role to control the signature of bulk KR field 
in our visible universe. However, the dynamics of $\chi^{(0)}(t,\varphi)$ are coupled to the evolution of the radion field $\Psi(t)$ and thus we need to obtain $\Psi(t)$ in 
order to determine the cosmological evolution of the KK wave function.\\

By eqn.~(\ref{energy densityII}), there remain two independent equations to fix the evolution of $\Psi(t)$ and $b(t)$, 
\begin{eqnarray}
 H_b^2 = \frac{1}{3}\bigg[\frac{1}{2}\dot{\Psi}^2 + \frac{20}{\sqrt{3}}\frac{\alpha k^5}{M^6}v_v^2 \Psi^4 \big[F\Psi^\omega - 1\big]^2\bigg] 
 + \frac{1}{3}\frac{\Omega_0}{b^6}\ ,
 \label{independent equation1II}
\end{eqnarray}
and
\begin{eqnarray}
 \ddot{\Psi} + 3H_b\dot{\Psi} + \frac{80}{\sqrt{3}}\frac{\alpha k^5}{M^6}v_v^2 \Psi^3 \big[F\Psi^\omega - 1\big]^2 = 0\ .
 \label{independent equation2II}
\end{eqnarray}
As mentioned above, 
we are interested to solve the equations at the early universe where the potential energy of the radion field is considered to be greater than 
the kinetic term (slow-roll approximation) i.e
\begin{eqnarray}
U_{rad}(\Psi) \gg \frac{1}{2}\dot{\Psi}^2\ .
 \label{slow roll approximationII}
\end{eqnarray}
By this approximation, eqns.~(\ref{independent equation1II}) and (\ref{independent equation2II}) become
\begin{eqnarray}
 H_b^2 = \frac{20}{3\sqrt{3}}\frac{\alpha k^5}{M^6}v_v^2 \Psi^4 \big[F\Psi^\omega - 1\big]^2 + \frac{1}{3}\frac{\Omega_0}{b^6}\ ,
 \label{slow roll equation1II}
\end{eqnarray}
and
\begin{eqnarray}
 3H_b\dot{\Psi} + \frac{80}{\sqrt{3}}\frac{\alpha k^5}{M^6}v_v^2 \Psi^3 \big[F\Psi^\omega - 1\big]^2 = 0\ .
 \label{slow roll equation2II}
\end{eqnarray}
Then, solving the above two equations for $\Psi(t)$ and $b(t)$, we get
\begin{eqnarray}
 \Psi(t) = \frac{\Psi_0}{\bigg[F\Psi_0^{\omega}-\big(F\Psi_0^{\omega}-\frac{\sqrt{\Omega_0}}{b_0^3\xi_0^2}-1\big)\exp
 {\big(-8\omega v_v\sqrt{\frac{5}{3\sqrt{3}}\frac{\alpha k^5}{M^6}}(t-t_0)\big)}\bigg]^{1/\omega}}\ ,
 \label{sol of radionII}
\end{eqnarray}
and
\begin{eqnarray}
 b(t) = C \bigg[1 + \sqrt{3\Omega_0}(t-t_0)\bigg]^{1/3} \exp{\bigg[2v_v\sqrt{\frac{5}{3\sqrt{3}}\frac{\alpha k^5}{M^6}} \big(g_1(t)-g_2(t)\big)\bigg]}\ .
 \label{sol of scaleII}
\end{eqnarray}
Recall that $F = \frac{1}{v_vf^{\omega}}\big(v_h - \frac{\kappa v_h^2}{2\sqrt{3}} + \frac{\kappa v_vv_h}{2\sqrt{3}}\big)$ and $\Psi_0$, $C$ are integration constants 
with $b_0=C\exp{[-\Psi_0^2/8]}$. Furthermore, $g_1(t)$ has the following expression: 
\begin{eqnarray}
 &g_1(t)&= - \frac{F\Psi_0^{\omega}}{(F\Psi_0^{\omega}-1)} 
 \frac{1}{16\omega v_v\sqrt{\frac{5}{3\sqrt{3}}\frac{\alpha k^5}{M^6}}}\Psi_0^2\nonumber\\ 
 &\times &2F_1\bigg(1,1,2+\frac{2}{\omega},\frac{F\Psi_0^{\omega}}{F\Psi_0^{\omega}-1}
 \exp{\big(8\omega v_v\sqrt{\frac{5}{3\sqrt{3}}\frac{\alpha k^5}{M^6}}(t-t_0)\big)}\bigg)\nonumber\\
 &\times &\exp{\bigg(8\omega v_v\sqrt{\frac{5}{3\sqrt{3}}\frac{\alpha k^5}{M^6}}(t-t_0)\bigg)} 
 \bigg(F\Psi_0^{\omega}-(F\Psi_0^{\omega}-1)\nonumber\\
 &\times &\exp{\big(-8\omega v_v\sqrt{\frac{5}{3\sqrt{3}}\frac{\alpha k^5}{M^6}}(t-t_0)\big)}\bigg)^{-2/\omega}\ ,
 \label{g1II}
\end{eqnarray}
where $2F1$ refers to an hypergeometric function. On the other hand, $g_2(t)$ is given by,
\begin{eqnarray}
 &g_2(t)&= - \frac{\Psi_0^{\omega}}{(F\Psi_0^{\omega}-1)} \frac{1}{16\omega v_v\sqrt{\frac{5}{3\sqrt{3}}\frac{\alpha k^5}{M^6}}}\Psi_0^2*\nonumber\\ 
 &\times &2F_1\bigg(1,1,1+\frac{2}{\omega},\frac{F\Psi_0^{\omega}}{F\Psi_0^{\omega}-1}
 \exp{\big(8\omega v_v\sqrt{\frac{5}{3\sqrt{3}}\frac{\alpha k^5}{M^6}}(t-t_0)\big)}\bigg)\nonumber\\
 &\times &\exp{\bigg(8\omega v_v\sqrt{\frac{5}{3\sqrt{3}}\frac{\alpha k^5}{M^6}}(t-t_0)\bigg)} \bigg(F\Psi_0^{\omega}-(F\Psi_0^{\omega}-1)\nonumber\\
 &\times &\exp{\big(-8\omega v_v\sqrt{\frac{5}{3\sqrt{3}}\frac{\alpha k^5}{M^6}}(t-t_0)\big)}\bigg)^{1-2/\omega}\ .
 \label{g2II}
\end{eqnarray}
Note that for $\alpha\rightarrow 0$, the solution of radion field and Hubble parameter become 
$\Psi(t) = \frac{\Psi_0}{\big[1 + \frac{\sqrt{\Omega_0}}{b_0^3\xi_0^2}\big]^{1/\omega}}=\Psi(t_0)$ and $H \propto \frac{1}{b^3}$ 
respectively. However, this result is expected since without any higher order curvature term (i.e $\alpha = 0$), the radion potential vanishes and  
 the radion field becomes constant $\Psi(t)=\Psi(t_0)$ while the Hubble parameter $\propto 1/b^3$ (solely due to the Kalb-Ramond field 
has an equation of state parameter $w=1$). Moreover, for $\Omega_0 = 0$, the solution turn out the one for pure $F(R)=R+\alpha R^2$ gravity in the Randall-Sundrum model, 
as found in \cite{tp_inflation}.\\

\begin{figure}[!h]
\begin{center}
 \centering
 \includegraphics[width=3.5in,height=2.5in]{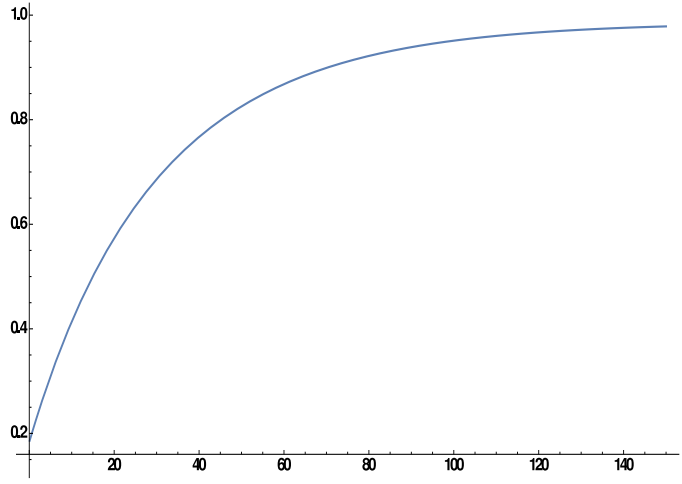}
 \caption{$\frac{T(t)}{<T>}$ vs $\tilde{t}$ for $\kappa v_v = \frac{\sqrt{\Omega_0}}{M^2} \simeq 10^{-7}$, $\frac{m_{\Phi}}{k} = 0.2$ and $\Psi_0 = 36$ (in Planckian unit).}
 \label{plot_brane_separation}
\end{center}
\end{figure}

Eqn.~(\ref{sol of radionII}) shows that the radion field decreases with the cosmic time and finally its bulk leads to 
(see eqn.(\ref{vev.radion.F(R)II})) asymptotically i.e
\begin{eqnarray}
 \Psi(t\gg t_0) = f\bigg[\frac{v_v}{v_h}\bigg]^{1/\omega} = <\Psi>\ .
 \nonumber
\end{eqnarray}
Thereby, the dynamics of the interbrane separation ($T(t)$) are as follows : $T(t)$ increases (as $\Psi(t)\propto e^{-k\pi T(t)}$) 
with the expansion of the universe and gradually goes to a stable value 
($k\pi <T> = \frac{4k^2}{m_{\Phi}^2}[\ln{(\frac{v_h}{v_v})} - \frac{\kappa v_v}{2\sqrt{3}}(\frac{v_h}{v_v} - 1)]$) asymptotically, as shown in Fig.~\ref{plot_brane_separation}.

Once we obtain the solution for $\Psi(t)$, we can obtain the evolution of the extra dimensional KR wave function $\chi^{(0)}(t,\varphi)$. 
Nevertheless, let us study whether the solution of the scale factor (\ref{sol of scaleII}) corresponds to an inflationary stage.

\subsection{Beginning of inflation}

In order to check whether the solution of the scale factor is consistent with an early inflationary stage, we expand $b(t)$ in the form 
of Taylor series (about $t=t_0$) and keep the terms up to the linear order in $t-t_0$:
\begin{eqnarray}
 b(t\gtrsim t_0)=b_0 \bigg[1 + \sqrt{3\Omega_0}(t-t_0)\bigg]^{1/3} 
 \exp{\bigg[2(F\Psi_0^{\omega}-1) \Psi_0^2 v_v\sqrt{\frac{5}{3\sqrt{3}}\frac{\alpha k^5}{M^6}}(t-t_0)\bigg]}\ ,
 \label{limiting scale factorII}
\end{eqnarray}
where $b_0$ is the value of the scale factor at $t=t_0$ and is related to the integration constant $C$ as,
\begin{equation}
 b_0 = C \exp{[-\Psi_0^2/8]}\ .
 \nonumber
\end{equation}
Eqn.~(\ref{limiting scale factorII}) leads to an accelerating expansion at $t \rightarrow t_0$ as follows:
\begin{eqnarray}
 \frac{\ddot{a}}{a}(t\gtrsim t_0&)&= \bigg[2(F\Psi_0^{\omega}-1) \Psi_0^2 v_v\sqrt{\frac{5}{3\sqrt{3}}\frac{\alpha k^5}{M^6}} 
 + \sqrt{\Omega_0}\big(1+\frac{1}{\sqrt{3}}\big)\bigg]\nonumber\\ 
 &\bigg[&2(F\Psi_0^{\omega}-1) \Psi_0^2 v_v\sqrt{\frac{5}{3\sqrt{3}}\frac{\alpha k^5}{M^6}} 
 - \sqrt{\Omega_0}\big(1-\frac{1}{\sqrt{3}}\big)\bigg]\ .
 \label{limiting accelerationII}
\end{eqnarray}
Note that under the condition
\begin{eqnarray}
2(F\Psi_0^{\omega}-1) \Psi_0^2 v_v\sqrt{\frac{5}{3\sqrt{3}}\frac{\alpha k^5}{M^6}} > \sqrt{\Omega_0}\big(1-\frac{1}{\sqrt{3}}\big)\ ,
\label{conditionII}
\end{eqnarray}
the early universe expands with an accelerating phase. Otherwise, the acceleration $\frac{\ddot{b}}{b}(t\rightarrow t_0)$ turns out negative. 
Recall that the on-brane KR field energy density ($\Omega_{KR}$) is proportional to $1/b^6$ as given by eqn.~(\ref{energy densityII})). Thereby, 
due to the inflationary expansion of the scale factor, $\Omega_{KR}$ rapidly decreases during the very early universe. However, 
eqn.~(\ref{limiting accelerationII}) clearly reveals that for $\alpha\rightarrow 0$, $\frac{\ddot{b}}{b}(t\gtrsim t_0)$ becomes less than zero 
i.e the early universe passes through a decelerating phase - solely due to the KR field having equation of state parameter $w=1$. 
Therefore, besides stabilizing the interbrane separation, the higher order curvature 
term also ensures the early inflationary stage subjected to the condition (\ref{conditionII}), which in turn provides a rapid decrease
of the Kalb-Ramond field energy density on the visible universe.

\subsection{End of inflation}

After obtaining the inflationary solution, it is important to evaluate whether the inflationary phase has a graceful exit in a finite time, 
as is connected to the resolution of the Horizon problem. The end of inflation can be defined as,
\begin{eqnarray}
 \frac{\ddot{b}}{b} = \dot{H}_b + H_b^2 = 0\ .        
 \label{end of inflationII}
\end{eqnarray}
Now let us estimate the time interval consistent with this condition. However, at the end of inflation, the term proportional to $1/b^6$ can be safely 
ignored and thus eqn.~(\ref{independent equation1II}) becomes,
\begin{eqnarray}
 H_b^2 = \frac{20}{3\sqrt{3}}\frac{\alpha k^5}{M^6}v_v^2 \Psi^4 \big[F\Psi^\omega - 1\big]^2\ .
 \nonumber
\end{eqnarray}
Differentiating both sides of this expression, one gets
\begin{eqnarray}
 \dot{H}_b = -\frac{160}{3\sqrt{3}}\frac{\alpha k^5}{M^6}v_v^2 \Psi^2\big(F\Psi^{\omega} - 1\bigg)^2\ .
 \label{time derivative of hubbleII}
\end{eqnarray}
Using the above expressions of $H_b^2$ and $\dot{H}_b$ in the eqn.~(\ref{end of inflationII}), we finally get the following 
condition on the radion field,
\begin{equation}
 \Psi = 2\sqrt{2} = \Psi_f = \Psi(t_f)\ ,
 \label{end value of radion fieldII}
\end{equation}
where $t_f$ is the time when the radion field takes the value $2\sqrt{2}$ (in Planckian units). Therefore, eqn.~(\ref{end value of radion fieldII}) 
clearly indicates that the inflationary era continues as long as the value of the radion field remains greater than $2\sqrt{2}$ (in Planckian units). 
With this information, one can determine the duration of inflation $t_f-t_0$ from the solution of $\Psi(t)$ (see eqn.~(\ref{sol of radionII})) as, 
\begin{eqnarray}
 \Psi(t_f) = \frac{\Psi_0}{\bigg[F\Psi_0^{\omega}-\big(F\Psi_0^{\omega}-\frac{\sqrt{\Omega_0}}{b_0^3\xi_0^2}-1\big)\exp
 {\big(-8\omega v_v\sqrt{\frac{5}{3\sqrt{3}}\frac{\alpha k^5}{M^6}}(t_f-t_0)\big)}\bigg]^{1/\omega}}\ .
 \nonumber
\end{eqnarray}
By simplifying, we get the expression of $t_f-t_0$ as follows,
\begin{eqnarray}
 t_f-t_0 = \frac{1}{8\omega v_v\sqrt{\frac{5}{3\sqrt{3}}\frac{\alpha k^5}{M^6}}}
 \ln\bigg[\frac{F\Psi_0^{\omega} - 1 - \frac{\sqrt{\Omega_0}}{b_0^3\Psi_0^2}}
 {F\Psi_0^{\omega} - \frac{\Psi_0^{\omega}}{\Psi_f^{\omega}}}\bigg]\ .
 \label{durationII}
\end{eqnarray}
Recall that $F = \frac{1}{v_vf^{\omega}}\big(v_h - \frac{\kappa v_h^2}{2\sqrt{3}} + \frac{\kappa v_vv_h}{2\sqrt{3}}\big)$ 
and $\omega = \frac{m_{\Phi}^2}{4k^2}$ with $m_{\Phi}^2$ given in eqn.~(\ref{mass_phiII}). Hence, the duration of inflation 
depends on the parameters $\alpha$ and $\Omega_0$, i.e. on the strength of the higher order curvature term and on the energy density of the KR field respectively. 
Therefore, in order to estimate $t_f-t_0$ explicitly, we need to determine the value of 
these parameters which, on the other hand, should be consistent with the observational constraints. 

\subsection{Spectral index, tensor to scalar ratio and number of e-foldings}

\begin{figure}[!h]
\begin{center}
 \centering
 \includegraphics[width=3.5in,height=2.0in]{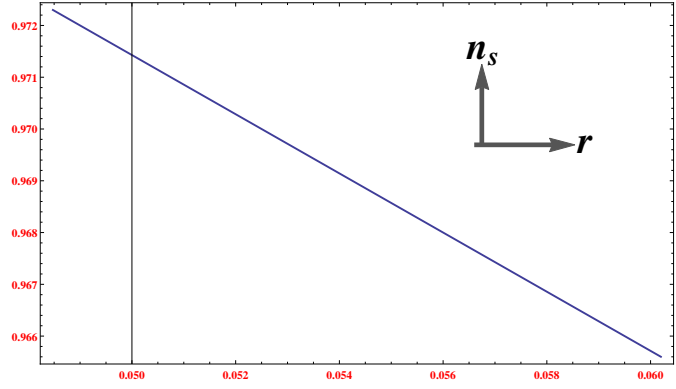}
 \caption{$n_s$ vs $r$ for $34<\Psi_0<38$ (in Planckian unit)}
 \label{observable_5d}
\end{center}
\end{figure}

As shown in previous sections, the results of Planck, 2018 \cite{Planck} put a certain constraint on the spectral index $n_s$ and the tensor to scalar ratio $r$ as 
$n_s = 0.9650 \pm 0.00661$ and $r < 0.07$ (combined with BICEP2/Keck - Array) respectively. As shown in Appendix-II, KR tensor $H_{\mu\nu\alpha}^{(0)}$ can be 
mapped to a derivative of a massless scalar field and thus $n_s$, $r$ are defined as follows (in terms of a 
dimensionless parameter $\epsilon_b = -\frac{\dot{H}_b}{H_b^2}$) \cite{4,5}:
\begin{eqnarray}
 n_s&=&1 - 6\epsilon_b\bigg|_{t=t_0} - 2\frac{\dot{\epsilon_b}}{H_b\epsilon_b}\bigg|_{t=t_0}\ ,\nonumber\\
 r&=&16\epsilon\bigg|_{t=t_0}\ .
 \label{definition}
\end{eqnarray}
Thereby, in order to scan the possible values of $\alpha$ and $\Omega_0$ provided by the constraints of Planck 2015, firstly we need to determine 
$\epsilon_b$ which determines the spectral index and the tensor to scalar ratio. For this purpose, we use the field equation 
$H_b^2 = \frac{1}{3}U_{rad}(\Psi) + \frac{\Omega_0}{6b^6}$. Differentiating both sides of this equation with respect to time, we get
\begin{eqnarray}
 2\dot{H_b} = -\frac{1}{9H_b^2}\bigg(\frac{\partial U_{rad}}{\partial\Psi}\bigg)^2 - \frac{\Omega_0}{b^6}\ ,
 \nonumber
\end{eqnarray}
where we have used the field equation for the radion field. These expressions for $\dot{H_b}$ and $H_b^2$ lead to the slow roll parameter $\epsilon_b$ as follows,
\begin{eqnarray}
 \epsilon_b = \frac{1}{2}\bigg[\frac{16p^2v_v^4\xi^6(F\Psi^{\omega}-1)^4 + \frac{3\Omega_0}{b^6}\bigg(pv_v^2\Psi^4(F\Psi^{\omega}-1)^2
 +\frac{\Omega_0}{2b^6}\bigg)}
 {\bigg(pv_v^2\Psi^4(F\Psi^{\omega}-1)^2 + \frac{\Omega_0}{2b^6}\bigg)^2}\bigg]\ ,\nonumber\\
 \label{slow roll parameter2}
\end{eqnarray}
where $p = \frac{k^3}{144M^6}$ and  $F = \frac{1}{v_vf^{\omega}}\big(v_h - \frac{\kappa v_h^2}{2\sqrt{3}} + \frac{\kappa v_vv_h}{2\sqrt{3}}\big)$.\\
By using the expression of $\epsilon_b$ along with eqn.~(\ref{definition}), $r$ and $n_s$ turn out to be,
\begin{eqnarray}
 r = 8\bigg[\frac{16p^2v_v^4\xi_0^6(F\Psi_0^{\omega}-1)^4 + \frac{3\Omega_0}{b_0^6}\bigg(pv_v^2\Psi_0^4(F\Psi_0^{\omega}-1)^2
 +\frac{\Omega_0}{2b_0^6}\bigg)}
 {\bigg(pv_v^2\Psi_0^4(F\Psi_0^{\omega}-1)^2 + \frac{\Omega_0}{2b_0^6}\bigg)^2}\bigg]\ ,\nonumber\\
 \label{ratio}
\end{eqnarray}
and
\begin{eqnarray}
 n_s = 1 - \frac{U_1}{U_2}\ ,
 \label{spectra index}
\end{eqnarray}
where $U_1$ and $U_2$ have the following expressions:
\begin{eqnarray}
 U_1&=&\bigg[384p^3v_v^6\Psi_0^8(F\Psi_0^{\omega}-1)^6 
 + \frac{18\Omega_0}{b_0^6}\bigg(pv_v^2\Psi_0^4(F\Psi_0^{\omega}-1)^2 + \frac{\Omega_0}{2b_0^6}\bigg)^2\nonumber\\ 
 &-&\frac{6\Omega_0}{b_0^6}\bigg(16p^2v_v^4\Psi_0^6(F\Psi_0^{\omega}-1)^4 + \frac{3\Omega_0}{b_0^6}
 \bigg(pv_v^2\Psi_0^4(F\Psi_0^{\omega}-1)^2 + \frac{\Omega_0}{2b_0^6}\bigg)\bigg)\nonumber\\
 &-&\frac{144\Omega_0}{b_0^6}p^2v_v^4\Psi_0^6(F\Psi_0^{\omega}-1)^4\bigg]\ ,
 \nonumber
\end{eqnarray}
and
\begin{eqnarray}
 U_2&=&\bigg(pv_v^2\Psi_0^4(F\Psi_0^{\omega}-1)^2 + \frac{\Omega_0}{2b_0^6}\bigg)\bigg(16p^2v_v^4\Psi_0^6(F\Psi_0^{\omega}-1)^4\nonumber\\
 &+&\frac{3\Omega_0}{b_0^6}\bigg(pv_v^2\Psi_0^4(F\Psi_0^{\omega}-1)^2 + \frac{\Omega_0}{2b_0^6}\bigg)\bigg)\ .
 \nonumber
\end{eqnarray}
As expected, the spectral index and the tensor to scalar ratio depend on the parameters $v_v$, $\Omega_0$ and $\Psi_0$. 
To fix these parameters, we use the observational results from Planck $2018$ \cite{Planck}. Here we take,
\begin{eqnarray}
\kappa v_v = \frac{\sqrt{\Omega_0}}{M^2} \simeq 10^{-7}\ .
\nonumber
\end{eqnarray}
These values of $v_v$ and $\Omega_0$ are consistent with the condition that is necessary for neglecting the 
backreaction of the bulk scalar field and the KR field on the background five dimensional spacetime. Then, by using eqns.~(\ref{spectra index}) and (\ref{ratio}) along 
with the values of $v_v$ and $\Omega_0$, the parametric plot for $n_s$ vs. $r$ is depicted in Fig.~\ref{observable_5d}, which 
clearly shows that within the interval $34<\Psi_0<38$ (in Planckian unit), 
both observable quantities $n_s$ and $r$ satisfy  the constraints provided by Planck 2018 \cite{Planck}. Furthermore, with the estimated values of $v_v$, $\Omega_0$ 
and $\Psi_0$, the duration of inflation $t_f-t_0$ becomes $10^{-10}$(Gev)$^{-1}$ as far as the ratio $m_{\Phi}/k$ (bulk scalar field mass to bulk curvature ratio) is taken to be $0.2$ \cite{GW}. 
This ratio of $m_{\Phi}/k$ leads to the stabilized interbrane separation as $k\pi<T> \simeq 36$ - required for solving the gauge hierarchy problem \cite{GW}. 
We also determine the number of 
e-foldings, defined by $N = \int_{0}^{\vartriangle t}H dt$ ($\vartriangle t = t_f-t_0$, duration of inflation), numerically and leading to $N \simeq 58$ (with $\xi_0 = 36$, in Planckian unit).
In table \ref{Table-2}, the results are summarised.

\begin{table}[!h]
 \centering
  \begin{tabular}{|c| c|}
   \hline \hline
   Parameters & Estimated values\\
   \hline
   $n_s$ & $0.9695$\\ 
   $r$ & 0.053\\
   $t_f-t_0$ & $10^{-10}$(GeV)$^{-1}$\\
   $N$ & 58\\
   \hline
  \end{tabular}%
  \caption{Estimated values of various quantities for $\kappa v_v = \frac{\sqrt{\Omega_0}}{M^2} \simeq 10^{-7}$ 
  and $\Psi_0 = 36.5$ (in Planckian unit).}
  \label{Table-2}
 \end{table}

 Table \ref{Table-2} clearly indicates that the present model may well explain the inflationary scenario of the universe 
 in terms of the observable quantities $n_s$ and $r$. Moreover, from Tables \ref{Table-1} and \ref{Table-2}, $n_s$ lies more closer 
 to the observational mean value ($<n_s>= 0.9650$) in four dimensions in comparison to the five dimensional Randall-Sundrum scenario.\\

\begin{figure}[!h]
\begin{center}
 \centering
 \includegraphics[width=3.5in,height=2.0in]{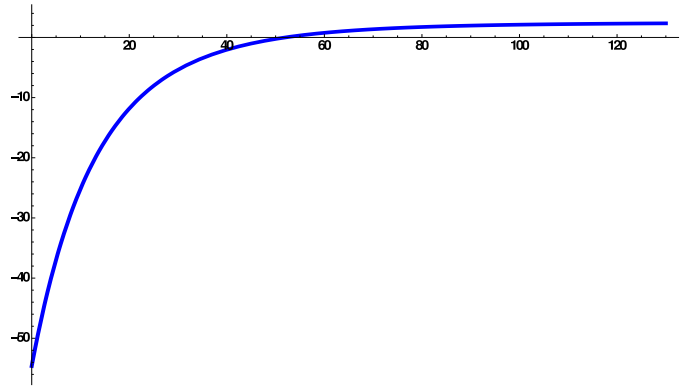}
 \caption{$q(t)$ vs $\tilde{t}$ for $\kappa v_v = \frac{\sqrt{\Omega_0}}{M^2} \simeq 10^{-7}$ and $\Psi_0 = 36$ (in Planckian unit).}
 \label{plot_deceleration_parameterII}
\end{center}
\end{figure}

 By using the solution of $b(t)$ (\ref{sol of scaleII}) along with the estimated values 
 of the parameters ($v_v$, $\Omega_0$, $\Psi_0$), the deceleration 
 parameter is depicted in \ref{plot_deceleration_parameterII}] in terms of the time variable $\tilde{t} = \frac{t}{t_f}N$, which shows that the early universe starts from 
 an accelerating stage with a graceful exit in a finite time.\\

  \subsection{Solution for the Kalb-Ramond extra dimensional wave function}

\begin{figure}[!h]
\begin{center}
 \centering
 \includegraphics[width=3.5in,height=2.5in]{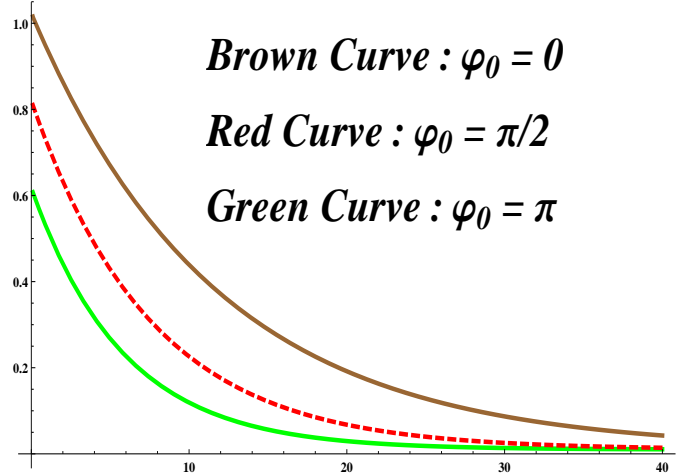}
 \caption{$\chi^{(0)}_{\varphi_0}(t,\varphi)$ vs $\tilde{t}$ for $\gamma = 0.15$, $\kappa v_v = \frac{\sqrt{\Omega_0}}{M^2} \simeq 10^{-7}$, $\frac{m_{\Phi}}{k} = 0.2$ and 
 $\Psi_0 = 36$ (in Planckian unit).}
 \label{plot_KR_wave_function2}
\end{center}
\end{figure}

The equation for the zeroth mode of the KR wave function $\chi^{(0)}(t,\varphi)$ follows from (\ref{wave function equation}), leading to,
 \begin{eqnarray}
 \bigg(\frac{\partial\chi^{(0)}}{\partial t}\bigg)^2 
 - \frac{1}{T^2(t)}e^{-2kT(t)\varphi}\chi^{(0)}\frac{\partial^2\chi^{(0)}}{\partial\varphi^2} = 0\ .
 \label{zeroth mode wave function equation}
\end{eqnarray}
The dynamics of the interbrane separation controls the evolution of $\chi^{(0)}(t,\varphi)$. The overlap of $\chi^{(0)}(t,\varphi)$ with the 
brane $\varphi=\pi$ (i.e. $\chi^{(0)}(t,\pi)$) regulates the coupling strengths among the KR field and various Standard Model fields on the visible brane. 
These interaction terms play the key role to determine the observable signatures of the KR field in our universe, such that we are interested to solve 
eqn.~(\ref{zeroth mode wave function equation}) in the vicinity of $\varphi=\pi$ (i.e. near the visible brane). 
Near the regime of $\varphi \simeq \pi$, eqn.~(\ref{zeroth mode wave function equation}) can be written as,
 \begin{eqnarray}
 \bigg(\frac{\partial\chi_v^{(0)}}{\partial t}\bigg)^2 
 - \frac{1}{T^2(t)}e^{-2k\pi T(t)}\chi_v^{(0)}\frac{\partial^2\chi_v^{(0)}}{\partial\varphi^2} = 0\ ,
 \label{zeroth mode wave function equation1}
\end{eqnarray}
where $\chi_v^{(0)}$ denotes the KR wave function near the visible brane. Eqn.~(\ref{zeroth mode wave function equation1}) can be solved by using the 
method of separation of variables as $\chi_v^{(0)}(t,\varphi) = f_1(t)f_2({\varphi})$. By this expression, eqn.~(\ref{zeroth mode wave function equation1}) 
turns out to be,
\begin{eqnarray}
 T^2(t)e^{2k\pi T(t)} \frac{1}{f_1^2}\bigg(\frac{df_1}{dt}\bigg)^2 = \frac{1}{f_2}\frac{d^2f_2}{d\varphi^2}\ .
 \label{zeroth mode wave function equation2}
\end{eqnarray}
As the left and right hand sides of eqn.~(\ref{zeroth mode wave function equation2}) are functions of time and $\varphi$ respectively, 
both sides can be separately fixed to a constant as follows:
\begin{eqnarray}
 T^2(t)e^{2k\pi T(t)} \frac{1}{f_1^2}\bigg(\frac{df_1}{dt}\bigg)^2 = \gamma^2\ ,
 \label{separation equation1}
\end{eqnarray}
and
\begin{eqnarray}
 \frac{1}{f_2}\frac{d^2f_2}{d\varphi^2} = \gamma^2\ ,
 \label{separation equation2}
\end{eqnarray}
where $\gamma$ is the constant of separation. The solution for the eqn.~(\ref{separation equation2}) is given by $f_2(\varphi) = e^{-\gamma\varphi}$, while 
eqn.~(\ref{separation equation1}) is solved numerically. Thereby, the solution for $\chi_v^{(0)}(t,\varphi)$ is given by $\chi_v^{(0)}(t,\varphi) = e^{-\gamma\varphi}f_1(t)$. 
Similarly in the vicinity of a general $\varphi = $constant hypersurface within the bulk ( i.e $\varphi\simeq \varphi_0$ ), 
the solution of KR wave function is given by $\chi^{(0)}_{\varphi_0}(t,\varphi) = e^{-\gamma\varphi}f_{\varphi_0}(t)$ where $f_{\varphi_0}(t)$ satisfies the differential 
equation : $T^2(t)e^{2k\varphi_0 T(t)} \frac{1}{f_{\varphi_0}^2}\bigg(\frac{df_{\varphi_0}}{dt}\bigg)^2 = \gamma^2$ ( obviously $f_{\varphi_0 = \pi}(t) = f_1(t)$ ). 
The solution of $\chi^{(0)}_{\varphi_0}(t,\varphi)$ along with the evolution of brane separation ( i.e $T(t)$, see eqn.(\ref{sol of radionII})) leads to the numerical plot 
for the time evolution of the KR wave function on the $\varphi = \varphi_0$ hypersurface, which is depicted in Fig.~\ref{plot_KR_wave_function2} for several values of $\varphi_0$. 
Fig.~\ref{plot_KR_wave_function2} reveals that the zeroth mode of the KR wave function $\chi^{(0)}(t,\varphi)$ decreases 
with time in the whole five dimensional bulk i.e for $0 \leq \varphi \leq \pi$. However, for a fixed $t$, $\chi^{(0)}(t,\varphi)$ has different values 
(in Planckian units) on the hidden ( $\varphi_0 = 0$ ) and visible brane ( $\varphi_0 = \pi$ ) 
and such hierarchial nature of $\chi^{(0)}(t,\varphi)$ (between the two branes) is controlled by the constant $\gamma$.\\
For $T(t) = <T>$, the zeroth mode of KR wave function acquires a constant value throughout the bulk and given by 
\begin{eqnarray}
\chi^{(0)}(t,\varphi)\bigg|_{T=<T>} = \sqrt{\frac{k}{<T>}}e^{-k\pi<T>}\ ,
\label{constant}
\end{eqnarray}
where we have used the normalization condition as shown in eqn.~(\ref{wave function normalization}). This result is also in agreement with \cite{sengupta1}. Using the above 
expression of $\chi^{(0)}(t,\varphi)\bigg|_{T=<T>}$, we obtain the coupling 
strengths of Kalb-Ramond field with $U(1)$ gauge field and fermion field on the visible brane as follows \cite{sengupta1}:
\begin{eqnarray}
 \lambda_{KR-U(1)} = \frac{1}{M_p}e^{-k\pi<T>}\ ,
 \label{coupling1}
\end{eqnarray}
and
\begin{eqnarray}
 \lambda_{KR-fer} = \frac{1}{M_p}e^{-k\pi<T>}\ .
 \label{coupling2}
\end{eqnarray}
Here $M_p=\sqrt{M^3/k}$. For $k<T> \simeq 12$ (required for solving the gauge hierarchy problem), $e^{-k\pi<T>}$ 
becomes of the order $10^{-16}$. Thereby, eqns.~(\ref{coupling1}) and (\ref{coupling2}) clearly indicate that 
the interaction strengths of the KR field to the matter fields are heavily suppressed over the usual gravity-matter coupling strength $1/M_p$. This may well serve as an 
explanation about why the behaviour of the present universe at large scales is solely governed by gravity and carries practically no observable footprints of antisymmetric Kalb-Ramond field.\\

\section{Conclusions}
\label{conclusion}
We have here addressed the issue of the absence of any perceptible footprints of  rank-two antisymmetric tensor fields, ordinarily known as Kalb-Ramond fields, in the 
framework of higher-order curvature gravity, both in four- as in five-dimensional spacetimes. Since all other type of fields, those with scalar, fermion and vector degrees of 
freedom, are known to be present in our Universe, the question of the absence of KR fields arises naturally. 

We have started from a particular $F(R)$ model, the well known Starobinsky model \cite{Starobinsky:1980te},  $F(R)= R + \frac{R^2}{m^2}$,  in the presence 
of a second rank antisymmetric, KR field propagating in a four dimensional spacetime. In such an scenario, we have obtained  
the cosmological evolution of the KR field in a flat FLRW universe. Our results reveal that the higher-order curvature term causes 
a gradual suppression of the energy density $\rho_{KR}$ of the KR field, eventually leading to an imperceptible footprint in the present universe. 
However, the effect of the KR field might still play a significant role in the early universe. This has led us to study the evolution of the KR field starting at the 
very early universe, when inflation is supposed to occur. We have shown that inflation is reproduced in our model due to the presence of higher-order terms in the action, 
so that the early universe expands through an accelerating phase, as far as the condition (\ref{condition F(R)}) is satisfied. This condition arises owing 
to an interplay which 
takes place between the strength of the higher-order curvature terms and the KR field itself, which at the end establishes whether the universe will go through an inflationary stage. 
In order to test the model with the most recent data (2018 run) from the Planck survey, we have matched the theoretical values for the spectral index of curvature perturbation ($n_s$) 
and tensor to scalar ratio ($r$), which are defined in terms of the slow-roll parameters, with the values  coming from the Planck observations. 
By relying in these definitions, the expressions of $n_s$ and $r$ are explicitly obtained, what provides some suitable values for the remaining free parameters 
($h_0$, $\xi_0$), while keeping $n_s$ and $r$ within the confidence regions provided by Planck 2018 (see Table \ref{Table-1}). In addition, we have also obtained an upper bound 
for the energy density of the KR field during the early universe, as $\rho_{KR} \leq 10^{70}$ (GeV)$^{4}$ (see also Ref.~\cite{logR}). 

By contrast, we have proven that in absence of higher-order curvature terms, the KR field behaves as a stiff-like fluid and consequently does not support inflation. 
However, authors in Ref.~\cite{1808.04315} showed that a stable de-Sitter solution can be achieved in the context of antisymmetric tensor fields,
by introducing a non-minimal coupling between the Ricci scalar and the tensor field. On the other hand, in the present paper, 
we argue that the minimal prescription (in the presence of an antisymmetric tensor field) can also give rise to an inflationary era, but in the presence 
of quadratic-curvature gravity. On top of this, we have also considered cubic gravity, where we have shown that, in the presence of the KR field, the spectral index and the tensor 
to scalar ratio satisfy the observable constraints.  However, a successful model for inflation also requires a graceful exit from it, within a finite time with an enough number 
of e-foldings. Hence, it is important to further analyse whether  $R^3$ gravity (or  a more general $R^n$ gravity with $n\geq3$) together with the KR field is consistent with an 
inflationary model, having a graceful exit, which we expected to investigate in a future work.

Moreover, we have also considered the same $F(R)$ model in a five dimensional Randall-Sundrum warped geometry within a two 3-brane scenario. Such 
braneworld scenario requires the stabilization of the interbrane separation (known as modulus or radion), for which one needs a stable potential term for the radion field. 
Here, the higher-order curvature term $\alpha R^2$ generates 
such a stable radion potential, fulfilling the requirement of modulus stabilization, since the radion potential 
$U_{rad}(\Psi)$ vanishes as the parameter $\alpha$ goes to zero, which clearly indicates that $U_{rad}$ is generated entirely by the extra gravitational terms in the action. 
In such an scenario, the cosmological evolution of the KR field is obtained by using a four-dimensional effective theory. However, when the KR field is allowed to propagate 
along the extra dimension, an additional wave function ($\chi^{(0)}$ arises, besides the on-brane part $H^{(0)}_{\mu\nu\lambda}$), which obviously gets coupled to the extra 
dimensional modulus field. Furthermore, the overlap between $\chi^{(0)}$ and the visible brane determines the coupling strength of the KR field to other matter fields. 
These interaction terms play a key role in the evaluation of the possible observable effects of the KR field in the current universe.

Due to the presence of $U_{rad}(\Psi)$, the modulus field $T(t)$ becomes dynamical, since $T(t)$ increases with the cosmic time ($t$) and finally leads to an stable value 
asymptotically, as shown in Fig.~\ref{plot_brane_separation}. This dynamics of the radion field triggers such evolution of the extra dimensional KR wave function $\chi^{(0)}(t,\varphi)$ 
(recall that $\varphi$ is the extra dimensional coordinate), which decreases with time in the full five dimensional bulk, i.e. for $0 < \varphi < \pi$. Moreover, 
for $T(t)=<T>$, $\chi^{(0)}(t,\varphi)$ becomes constant throughout the bulk, as obtained in Eq.~(\ref{constant}). 

Consequently, we have obtained the strengths of the  couplings of the KR field to several matter fields in the present visible universe. With the result that 
 such interaction strengths come with a heavily suppressed factor over the usual gravity-matter coupling $1/M_p$, thus obtaining a remarkably natural explanation of the 
 absence of any observation of the antisymmetric Kalb-Ramond field at large scales in the current universe.
 
In addition, the on-brane part, the energy density of the KR field $\Omega_{KR}$ has been found to behave as $1/b^6$ (here $b(t)$ is the scale 
factor of the visible brane), which clearly indicates that $\Omega_{KR}$ decreases more rapidly 
in comparison to radiation and pressureless matter. However, similarly to the four-dimensional case, Eq.~(\ref{energy densityII}) 
also entails that $\Omega_{KR}$ is large and may play a significant role during the early universe. After exploring the dynamics 
of the KR field during the early universe, when the scale factor is small compared to the present one, we have found solutions for the scale factor consistent with an 
early inflationary stage of the universe. Note that, in the absence of the higher-order curvature term $\alpha R^2$, 
the radion field becomes constant while the Hubble parameter varies as $H_b \propto 1/b^3$. This was to be expected, because for $\alpha \rightarrow 0$, the radion 
potential tends to zero, and thus the radion field has no dynamics leading to a Hubble parameter that goes as $H_b \propto 1/b^3$ (solely due to the KR field 
having equation of state parameter = 1). Furthermore, the duration of inflation ($t_f - t_0$) 
is also obtained by Eq.~(\ref{durationII}), which reveals that the accelerating phase of the universe ends within a finite time. 
We have also determined the spectral index and tensor to scalar ratio in the present context and found the corresponding constraints on the free parameters when compared to the  Planck 2018 values.

\section*{Acknowledgments}

EE, SDO and DSCG acknowledge the support by MINECO (Spain), project FIS2016-76363-P, and by AGAUR (Catalonia, Spain) project 2017 SGR247. DSCG is also funded by the grant No.~IT956-16 
(Basque Government, Spain).
TP acknowledges the hospitality by ICE-CSIC/IEEC (Barcelona, Spain), where a part of this work was done during visit. This article is based upon work from CANTATA COST 
(European Cooperation in Science and Technology) action CA15117,  EU Framework Programme Horizon 2020.

\appendix
\section*{Appendix I}
\label{appendix-I}
The field equation for Kalb-Ramond field is given by,
\begin{eqnarray}
 \partial_{\mu}\bigg[\sqrt{-g}\tilde{H}^{\mu\nu\lambda}\bigg] = 0\ ,
 \label{app2 1}
\end{eqnarray}
where $g$ is the determinant of the on-brane metric. Using the FRW metric ansatz, one obtains $\sqrt{-g} = a^3(t)$, where $a(t)$ 
is the scale factor of the universe. Thus, eqn.(\ref{app2 1}) takes the following form,
 \begin{eqnarray}
 &\partial_{\mu}&\bigg[a^3(t)\tilde{H}^{\mu\nu\lambda}\bigg] = 0\ ,\nonumber\\
 \Rightarrow &\partial_{0}&\bigg[a^3(t)\tilde{H}^{0\nu\lambda}\bigg] + \partial_{1}\bigg[a^3(t)\tilde{H}^{1\nu\lambda}\bigg]\ ,\nonumber\\
 &\partial_{2}&\bigg[a^3(t)\tilde{H}^{2\nu\lambda}\bigg] + \partial_{3}\bigg[a^3(t)\tilde{H}^{3\nu\lambda}\bigg] = 0\ .
 \label{app2 2}
\end{eqnarray}
Here the greek indices $\nu$, $\lambda$ run from $0$ to $3$. 
\begin{itemize}
 \item For $\nu = 2$ and $\lambda = 3$, eqn.(\ref{app2 2}) becomes
 \begin{eqnarray}
  &\partial_{t}&\bigg[a^3(t)\tilde{H}^{023}\bigg] + \partial_{x}\bigg[a^3(t)\tilde{H}^{123}\bigg]\ ,\nonumber\\
 &\partial_{y}&\bigg[a^3(t)\tilde{H}^{223}\bigg] + \partial_{z}\bigg[a^3(t)\tilde{H}^{323}\bigg] = 0\ .
 \label{app2 3}
 \end{eqnarray}
Due to the antisymmetric nature of the KR field, the last two terms of the above equation identically vanish. Furthermore, 
from eqn.~(\ref{sol off einstein equation}), $\tilde{H}^{023} = 0$. As a result, only the second term of eqn.~(\ref{app2 3}) 
survives and leads to the information that the non-zero 
component of KR field ($\tilde{H}^{123}$) is independent of the coordinate $x$ i.e $\partial_{x}\bigg[\tilde{H}^{123}\bigg] = 0$.

\item For $\nu = 1$ and $\lambda = 3$, eqn.(\ref{app2 2}) becomes
\begin{eqnarray}
  &\partial_{t}&\bigg[a^3(t)\tilde{H}^{013}\bigg] + \partial_{x}\bigg[a^3(t)\tilde{H}^{113}\bigg]\ ,\nonumber\\
 &\partial_{y}&\bigg[a^3(t)\tilde{H}^{213}\bigg] + \partial_{z}\bigg[a^3(t)\tilde{H}^{313}\bigg] = 0\ .
 \label{app2 4}
 \end{eqnarray}
 Here the third term survives, which ensures that $\tilde{H}^{123}$ is independent of $y$.
 
 \item For $\nu = 1$ and $\lambda = 2$, eqn.(\ref{app2 2}) becomes
 \begin{eqnarray}
  &\partial_{t}&\bigg[a^3(t)\tilde{H}^{012}\bigg] + \partial_{x}\bigg[a^3(t)\tilde{H}^{112}\bigg]\ ,\nonumber\\
 &\partial_{y}&\bigg[a^3(t)\tilde{H}^{212}\bigg] + \partial_{z}\bigg[a^3(t)\tilde{H}^{312}\bigg] = 0\ ,
 \label{app2 5}
 \end{eqnarray}
 where the fourth term sustains and gives $\partial_{z}\bigg[\tilde{H}^{123}\bigg] = 0$.
 
\end{itemize}

Therefore it is clear that the non-zero component of the Kalb-Ramond field i.e $\tilde{H}^{123}$ depends only on the time ($t$) coordinate.

\appendix*
\section*{Appendix II}
\label{appendix-II}
Due to the antisymmetric nature, $\tilde{H}_{\mu\nu\alpha}$ has four independent components 
in four dimensions and thus it can be equivalently expressed as 
vector field as,
\begin{eqnarray}
 \tilde{H}_{\mu\nu\alpha} = \varepsilon_{\mu\nu\alpha\beta}\Upsilon^{\beta}\ ,
 \label{app1 1}
\end{eqnarray}
where $\varepsilon_{\mu\nu\alpha\beta}$ is the Levi-Civita symbol and $\Upsilon^{\beta}$ is a vector field propagating in four dimensional spacetime. The four 
components of $\Upsilon^{\beta}$ are connected with the independent components of $\tilde{H}_{\mu\nu\alpha}$ as follows,
\begin{eqnarray}
\tilde{H}_{012} = h_1 = \Upsilon^{3}\ ,~~~~~~~~~~~~~~~\tilde{H}_{013} = h_2 = -\Upsilon^2\ ,\nonumber\\
\tilde{H}_{023} = h_3 = \Upsilon^{1}\ ,~~~~~~~~~~~~~~~\tilde{H}_{123} = h_4 = -\Upsilon^0\ .
\label{app1 2}
 \end{eqnarray}
Here, we assume FLRW metric as the ansatz,
 \begin{eqnarray}
  ds^2 = -dt^2 + a^2(t)\big[dx^2 + dy^2 + dz^2\big]\ .
  \nonumber
 \end{eqnarray}
 By this metric, the off-diagonal Einstein's equations become,
 \begin{eqnarray}
  \Upsilon_3\Upsilon^2 = \Upsilon_3\Upsilon^1 = \Upsilon_2\Upsilon^1 = \Upsilon_0\Upsilon^3 = \Upsilon_0\Upsilon^2 = \Upsilon_0\Upsilon^1 = 0\ .
  \label{app1 3}
 \end{eqnarray}
  The above set of equations clearly indicate that only one component of $\Upsilon^{\beta}$ is non-zero which reduces the independent components 
 of $\tilde{H}_{\mu\nu\alpha}$ to $1$. Therefore, in a spatially flat FLRW metric in four dimensions, $\Upsilon^{\beta}$ can 
 be expressed as a derivative of a massless scalar field $Z(x^{\mu})$ (i.e $\Upsilon^{\beta} = \partial^{\beta}Z$), 
 which further relates the KR field tensor with the scalar field as follows
 \begin{eqnarray}
  \tilde{H}_{\mu\nu\alpha}&=&\varepsilon_{\mu\nu\alpha\beta}\Upsilon^{\beta}\nonumber\\
  &=&\varepsilon_{\mu\nu\alpha\beta}\partial^{\beta}Z\ ,
  \label{app1 4}
 \end{eqnarray}
 Due to the FLRW metric, the scalar field $Z$ is considered to be homogeneous in space and thus its equation of motion turns 
 out to be,
 \begin{eqnarray}
  \ddot{Z} + 3H\dot{Z} = 0\ ,
  \label{app1 5}
 \end{eqnarray}
  where $H$ is the Hubble parameter. Then, by solving the above equation, one obtains
 \begin{eqnarray}
  \frac{\partial Z}{\partial t} \propto \frac{1}{a^3} = \frac{d}{a^3}\ .
  \label{app1 6}
 \end{eqnarray}
 Here $d$ is a proportional constant. By this solution of $\frac{\partial Z}{\partial t}$, the diagonal Friedmann equations take the following form-
 \begin{eqnarray}
 H^2&=&\frac{\kappa^2}{3}\bigg[\frac{1}{2}\dot{\xi}^2 + \frac{m^2}{8\kappa^2}\big(1 - e^{\sqrt{\frac{2}{3}}\kappa\xi}\big)^2 
 + \frac{1}{2}\dot{Z}^2\bigg]\nonumber\\
 &=&\frac{\kappa^2}{3}\bigg[\frac{1}{2}\dot{\xi}^2 + \frac{m^2}{8\kappa^2}\big(1 - e^{\sqrt{\frac{2}{3}}\kappa\xi}\big)^2 
 + \frac{d}{2a^6}\bigg]\ ,
 \label{app1 7}
\end{eqnarray}
and
\begin{eqnarray}
 2\dot{H} + 3H^2 = -\kappa^2\bigg[\frac{1}{2}\dot{\xi}^2 
 - \frac{m^2}{8\kappa^2}\bigg(1 - e^{\sqrt{\frac{2}{3}}\kappa\xi}\bigg)^2 + \frac{1}{2}\dot{Z}^2\bigg]\nonumber\\
 = -\kappa^2\bigg[\frac{1}{2}\dot{\xi}^2 
 - \frac{m^2}{8\kappa^2}\bigg(1 - e^{\sqrt{\frac{2}{3}}\kappa\xi}\bigg)^2 + \frac{d}{2a^6}\bigg]\ .
 \label{app1 8}
\end{eqnarray}
Recall that $\xi(t)$ is the scalar field which arises from the higher order curvature degree of freedom. Furthermore, the field equation for $\xi(t)$ is given by,
\begin{eqnarray}
 \ddot{\xi} + 3H\dot{\xi} - \sqrt{\frac{2}{3}}\frac{m^2}{4\kappa}e^{\sqrt{\frac{2}{3}}\kappa\xi}\big(1 - e^{\sqrt{\frac{2}{3}}\kappa\xi}\big) = 0\ .
 \label{app1 9}
\end{eqnarray}
Note that the above equations match with the field equations obtained in Eqns.~(\ref{independent equation1}) and 
(\ref{independent equation2}), by identifying the constant $d$ with $h_0$. This leads to the argument that the two representations 
($\tilde{H}_{\mu\nu\alpha}$ is expressed/ is not expressed by a vector field) are equivalent at the level of equation of motion.


\end{document}